%% file: main.tex
\documentclass[1p]{elsarticle}
\usepackage{makeidx} % allows for indexgeneration
\usepackage{amsmath}
\usepackage{amssymb}
\usepackage{ntheorem}
\usepackage{amsfonts} 
\usepackage{graphicx}
\usepackage{enumitem}
\usepackage{color}
\usepackage[linesnumbered,ruled,vlined]{algorithm2e}
\usepackage{algpseudocode}
\usepackage{booktabs}
\usepackage[section]{placeins}
\usepackage{tabularx}
\usepackage{framed}
\usepackage{url}
\usepackage{multirow}
\usepackage{caption}
\usepackage{float}
\usepackage[table,xcdraw]{xcolor}
\restylefloat{table}

\makeatletter
\def\ps@pprintTitle{%
 \let\@oddhead\@empty
 \let\@evenhead\@empty
 \def\@oddfoot{\footnotesize \copyright 2025. This manuscript version is made available under the CC-BY-NC-ND 4.0.}%
 \let\@evenfoot\@oddfoot}
\makeatother

\begin{document}
\begin{frontmatter}
	
	\title{Optimisation of cyber insurance coverage with selection of cost effective security controls\footnote{doi: \url{https://doi.org/10.1016/j.cose.2020.102121}}.}
	
	\author[add2,add1]{Ganbayar Uuganbayar}
	\ead{ganbayar.uuganbayar@\{iit.cnr.it, unitn.it\}}
	\author[add2]{Artsiom Yautsiukhin}
	\ead{artsiom.yautsiukhin@iit.cnr.it}
	\author[add2]{Fabio Martinelli}
	\ead{fabio.martinelli@iit.cnr.it}
	\author[add1,add3]{Fabio Massacci}
	\ead{fabio.massacci@ieee.org}
	
	\address[add2]{Istituto di Informatica e Telematica, Consiglio Nazionale delle Ricerche, Pisa, Italy}
	
	\address[add1]{Department of Information Engineering and Computer Science (DISI), University of Trento, Italy}
	
	\address[add3]{Department of Computer Science, Vrije Universiteit, Netherlands}

	\begin{abstract}
		Nowadays, cyber threats are considered among the most dangerous risks by top management of enterprises. One way to deal with these risks is to insure them, but cyber insurance is still quite expensive. The insurance fee can be reduced if organisations improve their cyber security protection, i.e., reducing the insured risk. In other words, organisations need an investment strategy to decide the optimal amount of investments into cyber insurance and self-protection. 

		In this work, we propose an approach to help a risk-averse organisation to distribute its cyber security investments in a cost-efficient way. What makes our approach unique is that next to defining the amount of investments in cyber insurance and self-protection, our proposal also explicitly defines how these investments should be spent by selecting the most cost-efficient security controls. Moreover, we provide an exact algorithm for the control selection problem considering several threats at the same time and compare this algorithm with other approximate algorithmic solutions.
	\end{abstract}
	
	
\end{frontmatter}

	\input{01_Introduction.tex}
	\input{06_RelatedWork.tex}
	\input{02_Alt.tex}
	\input{03_SolutionTheory.tex}
	\input{05_CaseStudy}
	\input{Discussion.tex}
	\input{07_Conclusion.tex}

\bibliographystyle{elsarticle-num-names}

	\renewcommand{\thesubsection}{\Alph{subsection}}
	\input{99_Apendix}
\end{document}

%% file: 01_Introduction.tex
\section{Introduction}
Cyber security losses due to successful cyber attacks grow every year \cite{Symantec,Wanna}. This increase can be explained by a number of factors, such as increasing reliance of business and society on IT systems, fast growth of the number of interconnected end devices, lack of security awareness, maturation of the cyber crime world \cite{Cisco,HOMO}, etc. These losses have already spurred top management of companies to consider cyber security risks among the most significant ones. 

Typically, organisations have 4 basic options to treat risks: \emph{avoid} it by abandoning risky business, \emph{reduce} it by implementing security controls, \emph{transfer} it to another entity (e.g., buy insurance), \emph{accept} it if there is no other suitable choice \cite{OCTAVE,MAGERIT,NIST800}. The decision about treating cyber risks must be taken after careful analysis of the available options and ensuring that the selected strategy is cost-effective. Among the four risk treatment options, risk reduction and risk transferring require particular attention, because both of them require additional investments. Thus,  organisations need an instrument which helps them to identify the required investments in different risk treatment strategies to minimise the expected losses. 

Cyber insurance has received a lot of attention recently as more and more insurance companies enter the cyber market and their products become more mature. Also, numerous research papers devote significant attention to this topic \cite{Marotta2017,PartnerRe,Bohme-L,SoK2020}. For instance, \cite{Marotta2017,SoK2020} systematically reviewed the problems related to cyber insurance, where \cite{SoK2020} points out that the area needs more technical-oriented solutions to replace qualitative analysis. Some researchers further investigated the relation between cyber insurance and security investment. For instance, Massacci et al., \cite{Massacci} highlighted that cyber insurance might trigger the drop of security investment when attackers are fully informed about the security posture of an organisation. On the other hand, several works are devoted to making cyber insurance market profitable and encouraging organisation to invest in self-protection \cite{R.P.Majuca2006,Marotta2017}. The models used in the literature help to identify the most appropriate amount of investments in insurance and self-protection. These investment models simply assume that the probability of an attack is dependent on the amount of investment. In other words, the models do not tell how the organisation should spend these investments, i.e., how to select the required security configuration. Moreover, these models consider one threat (and one investment to probability dependency) while in reality organisation faces several cyber threats of a different kind which may cause different losses (e.g., see examples of threats from the ISO27005 standard \cite{ISO27005}).    

Risk reduction requires definition and implementation of a security configuration by the installation of various security controls to reduce the risk. Thus, an optimisation problem should be solved here, to select the most cost-efficient controls. For instance, in \cite{NIST800,Chand,Chand0}, the authors defined potential threats and controls in different systems, while an organisation may find it difficult to implement all required controls.

A typical mathematical problem to address the variety of choices is the family of knapsack problems \cite{knapsack,ANDREW1,Smeraldi,Maya}. A direct formulation for selection of cyber security configuration \cite{bazgan} has several limitations. First, the knapsack problem will always use as much of the budget as the limit allows (taking into account the discrete costs for security controls), even if the application of security controls provide fewer benefits (in the reduction of cyber risks) than their costs. Naturally, this approach is not the most cost-effective.

Second, usage of risk as the utility function makes the solution computationally ineffective (i.e., existing pseudo-polynomial solutions, like dynamic programming, cannot be applied). In some specific cases, if just one threat is considered (\cite{Symantec,ANTON} efficient solutions are possible, but for a comprehensive case the computation is not trivial. Therefore, many researchers apply approximate solutions (e.g., Genetic Algorithms (GA) \cite{LOREN,Gupta}), accepting the risk of receiving the only near-optimal answer. We look for a satisfying solution in the sense of Simon \cite{Simon}.

\subsection{Main contributions}
The core contribution of this paper is an approach for risk-averse organisations to determine security investments in various risk treatment options (in particular for risk reduction and cyber insurance) and determine specific security controls to be applied.
In particular, this article provides
\begin{itemize}
	\item a theoretical analysis of the distribution of the cost-efficient investments if cyber insurance is available,
	\item an explicit model to link investments with selected security controls, assuming that some initially selected controls could be not optimal,
	\item an exact solution for solving the optimisation problem (based on multi-objective knapsack problem),
	\item analysis of the proposed algorithms and approximate solutions (Greedy and GA), considering the effect of quantity and quality of inputs on the results.    
\end{itemize}

The theoretical analysis is based on the utility theory and assumes a competitive insurance market in place. The resulting optimisation problem is represented as a multi-objective knapsack problem where minimisation of risks for every threat is a separate objective. The algorithmic solutions for the problem are based on the dynamic programming \cite{bazgan}. Moreover, we enhance the algorithm by applying our projection idea and compare the results by conducting several experiments.

This paper is organised as follows. We start with the analysis of the related literature in Section~\ref{sec:RelatedWork}. Section~\ref{sec:Prob} is dedicated to the formal problem statement definition and detailed discussion on its applicability.  Section~\ref{sec:solutions} is devoted to the adaptation of the existing algorithmic solutions for our problem. Section~\ref{sec:CaseStudy} provides experimental results comparing the proposed algorithms with some approximate solutions (i.e., Greedy and Genetic Algorithms) and analyse their applicability. Section~\ref{sec:MAinDiscussion} systematically discusses the assumptions we consider in the paper and proposes further directions to extend our work. Finally, conclusion and achieved results are outlined in Section~\ref{sec:Discussion}. 

%% file: 06_RelatedWork.tex
\section{Related work}
\label{sec:RelatedWork}

To properly manage cyber risks, organisations should assess their risks \cite{FAB-17,Atif0} and define an efficient risk investment strategy for treating them \cite{NIST800,Chand}. In this work, we assume an organisation to be risk-averse and consider the option of \emph{insuring} some of its risks. To minimise the insurance premium, the organisation should invest in self-protection and \emph{optimise these security investments} by selecting the most cost-efficient security controls. In contrast to many existing papers, not only do we (prove and) formulate the problem for optimisation, but also propose our \emph{algorithmic solution} which is adapted for our concrete problem (rather than being generic) and returns the optimal (exact) answer.     
\subsection{Cyber insurance}
In most surveys and annual reports \cite{PartnerRe,PWC}, cyber insurance market is gradually growing and positively impacting on cyber security, although it faces some challenges \cite{Marotta2017,ENISA}. Also, Savino Dambra et al., \cite{SoK2020} underlined that current approaches for risk assessment are mostly qualitative and do not provide monetary information required for the cyber insurance underwriting process. The authors argue that the situation can be improved by applying data-driven methodologies and development of automatic tools. Scientific studies are largely focused on the problem of incentivising organisations to increase their cyber security investments if cyber insurance option is available \cite{Shetty2010,Schwartz2014,Lelarge2009,Ogut2004,R.P.Majuca2006}. Some researchers \cite{Ogut2004,Ehrlich1992,Massacci} claim that an organisation may invest less once the cyber insurance option becomes available. On the other hand, others \cite{R.P.Majuca2006,Lelarge2009}, i.e., Ruperto P.Majuca et al., found out that organisations are encouraged to invest for the self-protection to obtain a lower premium. Mukhopadhyay et al., \cite{MUKH} highlights that basic and concrete security investments for security controls are mapped to the affordable insurance premium. In contrast to the existing work on cyber insurance, we provide a cost-efficient approach for selecting concrete security controls and define how to map it with investment-probability correlation. To best of our knowledge, this problem has been often avoided by researchers and has not been investigated thoroughly.

\subsection{Optimisation of security configuration}

The problem of selecting the best security controls received a lot of attention from various researchers. For example, Chung et al., \cite{Chun} considered selecting the optimal security controls based on their Return on Investment (RoI) metric. This approach is similar to our Greedy algorithm and, as our experiments show, it often leads to only near-optimal answers.

Several authors \cite{Smeraldi,Sawik,DEWRI,leechoi} focused on formulating the problem of optimising best controls selection where they applied the basic knapsack problem \cite{knapsack}. Even though the main concept is similar, the authors assumed various limitations and modifications of the knapsack problem. In particular, T. Sawik \cite{Sawik} considered minimization of expected worst-case cost using Value at Risk (VaR) and Conditional Value-at-Risk (CVaR) values. Lee at al., \cite{yevseyeva1,yevseyeva2} considered the return on investment, expected losses and required resources as additional constraints. Dewri et al., \cite{DEWRI} and Viduto et al., \cite{VALEN,VALENTINA} formulated a problem considering both residual risk and cost of security controls as separate objectives to be minimised. Smeraldi and Malacaria \cite{Smeraldi} considered different assets to protect and binary interdependency of control. Fielder et al., \cite{ANDREW1} and Dewry et al., \cite{DEWRI} merged a knapsack representation of the problem with a game-theoretic approach to decide which strategy is better for protecting against an attacker. Rees et al., \cite{LOREN} considered uncertainty in the optimisation problem and used fuzzy values for risk computation.

Many of these formulated problems are considered as multi-objective problems \cite{Smeraldi,Sawik,DEWRI,VALEN}. We should underline that objectives considered by the authors (minimisation of losses, minimisation of costs, maximisation of return of investments, etc.) differ from our vision of objectives (i.e., minimising the risks caused by each threat). In this article, we prove that to optimise investments in self-protection and insurance we must consider total expenditure (i.e., minimise the sum of insured risk and total cost of controls). 

The core difference of our paper from these works is that we propose an exact algorithmic solution based on dynamic programming and adapt it for our problem, while the authors of the mentioned papers simply apply existing off the shelves solutions. For example, \cite{Sawik} applied mixed-integer programming approach to analyse a simple example with 10 threats and 10 controls (something, our algorithm can easily cope with). Smeraldi \cite{Smeraldi} considered classic dynamic programming and greedy algorithms. Other authors \cite{LOREN,DEWRI,leechoi,VALENTINA} used various evolutionary algorithms. Our experiments prove that GA is faster and reliable enough (with high settings), although it still may fail to produce the optimal answer. Viduto et al., \cite{VALEN} provided their solution called Multi-objective Tabu Search (MOTS), but their solution solves a slightly different problem which considers residual risk and security controls objectives separately.

\subsection{Algorithmic solutions for optimisation problems}

Dynamic programming \cite{MART,TOTH} is one of the main solutions for the knapsack problem as it finds the optimal answer through iterations until a certain condition meets. In the work of Bazgan et al., \cite{bazgan}, 0-1 multi-objective knapsack problem (considered 3 objectives for the experiment) has been solved by proposing dynamic programming, where the authors used several complementary dominance relations at different states and conducted experimental validation. In our paper, we improved the algorithm taking into account the peculiarities of our concrete problem (i.e., projection idea). On the other hand, evolutionary algorithms, i.e., Genetic Algorithm, are often considered in both literature and applications due to its capability of finding the optimal or near-optimal solutions in a much shorter time \cite{KHURI,WHIT}.

There is several attempts to improve the performance of GA by many researchers \cite{Maya,Gupta,VALENTINA,Ahmad,GORAN}. 
Maya Hristakeva and Dipti Shrestha \cite{Maya} proposed a GA-based solution for 0-1 knapsack problem and compared two selection methods, roulette-wheel and group selection (they devised a name for their approach). As an outcome, they claim that the group selection with elitism (copying some chromosomes without doing crossover technique) outperforms the roulette-wheel selection in different cases, i.e., increasing the number of population. Another work to improve the accuracy of GA, Gupta et al., \cite{Gupta} proposed a hybrid solution to create a better first population. They applied "fcheck" function to the initialization step to check whether the created population meets the criteria they set. The idea results in more good chromosomes in the population, which eventually improved the accuracy of the outcome. Also, Ahmad et al., \cite{Ahmad} proposed a linear regression analysis for creating the most efficient and fit population, yet this work is dedicated to another problem, travelling salesman problem (TSP). We have adapted some ideas of the aforementioned works to initialize the first population and improved them to fit into our work. 

Also, GA has been applied to various applications, including cyber security. For instance, Suhail Owais et al., \cite{SUHAIL} surveyed to apply GA to Intrusion Detection Systems (IDS) techniques and Goranin et al., \cite{GORAN} adapted GA to find the controls to mitigate the propagation of worms through the Internet. 

%% file: 02_Alt.tex
\section{Problem specification}
\label{sec:Prob}
In this section, we formalise our main problem gradually adding the required features. First, the optimisation problem for the security of a system is described, then the insurance option is added and the whole model is considered from the point of view of utility theory in the usual way for analysis in the insurance literature \cite{Marotta2017,Ogut2004,Ehrlich1992}. The main parameters used in this paper are listed in Table~\ref{tab:Notat}.

\begin{table*}[!t]
	\begin{center}    
		\resizebox{0.8\textwidth}{!}{
			\footnotesize
			\begin{tabular}{| c  l | c  l |}
				\hline
				$W^0$ & - initial wealth & $ \vec{\pi} $ & - probability of a threat \textit{survival}\\
				\hline
				$x$ & - security investment & $ \vec{F} $ & - expected number of threats \textit{attempts}     \\
				\hline
				$  c $ & - cost of a control & $\vec{p}$ & - probability of a threat \textit{occurrence}\\    
				\hline
				$P$ & - premium & $\vec{z}$ & - \textit{number} of threat occurrences \\
				\hline 
				$ K $ & - a set of \textit{available} controls & $ \vec{L} $ & - loss    \\
				\hline
				$ K_s $ & - a set of \textit{selected} controls & 
				$ \vec{I}$  & - indemnity     \\
				\hline
			\end{tabular}
		}
	\end{center}
	\caption{Notations adopted in this work}
	\label{tab:Notat}
\end{table*}

\subsection{No insurance case}
\label{sec:NoInsurance}
Consider an organisation which has conducted risk identification phase of risk assessment and has identified $ n_t~ (n_t \in \mathbb{N}^+) $ relevant threats. For each threat, a corresponding expected loss has been defined $ \vec{L} = \langle L^1, L^2, ..., L^{n_t} \rangle $, where $\vec{L}$ is a vector and $L^i (1 \leq i \leq n_t) $ is its i-th member. To decrease the amount of losses, the organisation invests $x$ in self-protection. This investment will be spent for installation and application of a set of security controls $K_s$, which can be seen as a subset of the set of all possible security controls $K$ (e.g., the ones that could be found in ISO27002 \cite{ISO} or NIST 800-53 \cite{NIST800}). Let the cost of a control be a function and its result be a finite non-negative value $c:K\mapsto \mathbb{N}^{+}$ (i.e.  thousands of Euro). The overall cost of installed controls $K_{s}\subseteq K$ ($c({K_{s}})$) is computed as: 
\begin{align} c({K_{s}}) = \sum_{\forall k\in K_{s}} c(k). \end{align}

The probability of threat $i$ successfully passing through all installed security controls is denoted as $p^{i}(K_s)$, which eventually can be seen as a vector for all $n_t$ threats, $ \vec{p}(K_s) = \langle p^1(K_s), ~p^2(K_s), ..., ~p^{n_t}(K_s) \rangle $. Now, if we know the frequency of threats occurrences $\vec{F} = \langle F^1, F^2, ..., F^{n_t} \rangle $, the overall risk for the organisation can be found as follows.
\begin{align}
Risk (K_s, x, \vec{L}) = (\vec{F}\odot\vec{p}(K_s))\times\vec{L},
\label{OverAllRisk}
\end{align}
where $\vec{a} \times \vec{b}$ is a usual matrix multiplication of two vectors given as $ \vec{a} \times \vec{b} = \sum_{i=1}^{n_t} a^i \cdot b^i $ and the Hadamard product of two vectors $ \vec{a} $ and $ \vec{b} $ is a vector $ \vec{c} = \vec{a} \odot \vec{b} = \langle a^{1}\cdot b^{1},a^{2}\cdot b^{2},...,a^{n_{t}}\cdot b^{n_{t}}\rangle $. In this paper, we omit the symbol for the transposition of vectors, required for proper representation of matrix multiplication, to simplify the formalisation since this precision is not crucial for the understanding of the paper. 

Since a threat may occur more times than once in an observed period, and harm the organisation more than once, we need to take into account the distribution of probabilities with respect to the number of threat occurrences. Let $ \vec{z} = \langle z^1, z^2, ..., z^{n_t} \rangle $ be a random vector of numbers of threat occurrences (one per threat) and $p(\vec{z}|K_{s})$ be the probability that the considered organisation will face $\vec{z}$ incidents for some period of time conditional on the implemented controls $K_{s}$. If the organisation has $W^0$ as an initial wealth (or expected benefit from its core business), the wealth after occurrence of $\vec{z}$ threats can be defined as:
\begin{align}
W(\vec{z},K_{s},x)=W^0-x-\vec{z}\times\vec{L} 
\end{align}
The goal of the organisation is to maximise its expected wealth, i.e., 
\begin{align}
E[W(\vec{z},K_{s},x)]=\sum_{\forall \vec{z}} (W^0-x-\vec{z}\times\vec{L})\cdot p(\vec{z}|K_{s})=\nonumber \\
W^0-x-\sum_{\forall \vec{z}}(\vec{z}\cdot p(\vec{z}|K_{s})\times\vec{L} 
\label{eq:WealthNoInsurance}
\end{align}
We note that $\sum_{\forall z}(\vec{z} \cdot p(\vec{z}|K_{s})$\footnote{"$ \cdot $" is the scalar multiplication defined as $\vec{a}\cdot b = \langle a^{1}\cdot b,a^{2}\cdot b,...,a^{n_{t}}\cdot b\rangle$} is the mean number of occurrences, previously defined as $\vec{F}\odot\vec{p}(K_s)$. Finally, our optimisation problem can be seen as
\begin{align}
&\max_{x, K_{s}} E[W(\vec{z},K_{s},x)]=W^0-x-(\vec{F}\odot\vec{p}(K_s))\times\vec{L}=W^0-x-Risk (K_s, x, \vec{L}) \\
&~~~or~ \nonumber\\
&\min_{x, K_{s}} [x+(\vec{F}\odot\vec{p}(K_s))\times\vec{L}]
\label{eq:MaxThirdFirst}
\end{align}

\subsection{Insurance case}
If an organisation buys insurance, it pays some premium $P$ regularly and expects some coverage of a loss if a threat occurs (called indemnity $ \vec{I}$). Indemnity is also a vector of size $n_t$, since depending on the purchased insurance product, different threats may get different coverage. The coverage is always lower than the loss itself, i.e., $\forall i,~I_i\leq L_i$. In cyber insurance, a premium is usually computed through the estimated risk. A usual assumption is to consider a competitive insurance market \cite{Marotta2017} for which the premium is equal to risk for the insurer (i.e., $P=Risk (K_s, x, \vec{I}))$. In case of purchased insurance, the resulting wealth for the organisation after occurrence of $\vec{z}$ threats (i.e., similar to Equation~\ref{eq:WealthNoInsurance}) can be found as:
\begin{align}
&W(\vec{z},K_{s},x,\vec{I})=W^0-(\vec{F}\odot\vec{p}(K_{s}))\times \vec{I}-x-\vec{z}\times(\vec{L}-\vec{I}), \\
&where~ \vec{I}-\vec{L}=\langle I^{1}-L^{1},I^{2}-L^{2},...,I^{n_{t}}-L^{n_{t}}\rangle. \nonumber
\end{align}

Similar to other economic models \cite{Marotta2017,Ogut2004,Ehrlich1992}, we assume an organisation to be risk-averse and use utility of possessing a certain amount of wealth $U(W)$ instead of the wealth $W$ itself. The utility function is considered to be continuous, non-decreasing, and concave, i.e., $U'(W)>0$ and $U''(W)<0$.
\begin{align}
&U(W(\vec{z},K_{s},x,\vec{I}))=U(W^0-(\vec{F}\odot\vec{p}(K_{s}))\times \vec{I}-x-\vec{z}\times(\vec{L}-\vec{I})).\label{eq:utility} 
\end{align} 
Finally, the expected utility is equal to:
\begin{align}
& E[U]= \sum_{\forall \vec{z}}p(\vec{z}|K_{s}) \cdot U(W^0-(\vec{F}\odot\vec{p}(K_{s}))\times \vec{I}-x+\vec{z}\times(\vec{I}-\vec{L})). \label{eq:ExpectedUtility} 
\end{align}
Our goal transforms into maximisation of the expected utility ($E[U]$) by selecting the optimal $x$, $\vec{I}$ and $K_{s}$. 

In these settings, it is possible to prove that the optimal $\vec{I}$ for an organisation is equal to $\vec{L}$ (see the proof in the Appendix~\ref{sec:Indemnity}). In other words, a risk-averse organisation insures all risks left after reduction if insurance option is available. This allows us to reduce Equation~\ref{eq:ExpectedUtility} to: 
\begin{align}
\max\limits_{x, K_{s}}U(W^0-x -(\vec{F}\odot \vec{p}(K_{s}))\times\vec{L}). \label{eq:MaxSecond}
\end{align}
Since the utility function is non-decreasing, we need to maximise its argument, or simply minimise the following part, which we call \emph{expenditure} in the following:
\begin{align}
\min\limits_{x, K_{s}} (x +(\vec{F}\odot \vec{p}(K_{s}))\times\vec{L}). \label{eq:MaxThird}
\end{align}

At this point we need to note, that our problem becomes equivalent to the risk-neutral approach, i.e., if $U(W)=W$ as in Section~\ref{sec:NoInsurance}, for selecting security budget distribution and controls selection if no insurance is available (in this case, the premium becomes the accepted residual risk). The organisation should simply minimise its investments in self-protection and residual risk. Thus, our further contribution could be applied if one of the conditions (Equations \ref{eq:MaxThirdFirst} and \ref{eq:MaxThird}) described above is found to be applicable.

\subsection{Selection of security controls}

Let $K_s|x$ denotes a set $K_s$ which minimises Equation \ref{eq:MaxThird} for some fixed investment $x$. To minimise Equation~\ref{eq:MaxThird} (as well as the minimisation condition in Equation~\ref{eq:MaxThirdFirst}), we need to compute $p(K_{s}|x)$ and determine the procedure for selection of $K_s$ in a way to minimise this component and ensure that we do this with investments less or equal to $x$. 

Let $ \pi^{j}(k) \in [0;1]$ be the probability that a threat $j$ passes through (survives) control $ k \in K_{s} $; control $k$ completely eliminates threat $j$ if $\pi^{j}(k) = 0$, and is entirely powerless against the threat if $ \pi^{j}(k) = 1 $. Let $\vec{\pi}(k)$ be a vector of all probabilities of survival if control $k$ is installed, and the overall probability of survival $\vec{p}({K_{s}}$) can be computed as\footnote{We assume effects of controls to be independent from each other.}:
\begin{align} \vec{p}({K_{s}}) = \prod_{\forall k\in K_{s}} \vec{\pi}(k),  \label{eq:OverallProb}
\end{align}
where $\prod_{\forall k\in K_{s}}$ stands for the Hadamard product.

Now, we say that  $\vec{p}(K_{s}|x)=\vec{p}({K_{s}})$ if $K_{s}$ minimises Equation~\ref{eq:MaxThird} and its overall cost is below $x$. Finally, the optimisation problem considered in this paper can be seen as:
\begin{align} 
\min\limits_{\forall K_{s}\subset K} (\vec{F}\odot \left[ \prod_{\forall k\in K_{s}} \vec{\pi}(k)\right])\times \vec{L}+x~~~and~~~
\sum_{\forall k\in K_{s}} c(k) \leq x \label{eq:Cond}.
\end{align}
Equation~\ref{eq:Cond} is the core problem we are tackling with in this paper.

\emph{Our goal in this article is to find the efficient distribution of investments, i.e., self-investments, insurance and accepted risk. Furthermore, we go further and explicitly show how these investments should be used by selecting the best set of security controls to achieve this efficient budget distribution.} Considering the efficient distribution of investments and selection of controls helps us to achieve the desired cost-effectiveness, in contrast to efficient threat mitigation only considered in many of the related works \cite{ANDREW1,Smeraldi,Sawik,DEWRI}.

%% file: 03_SolutionTheory.tex
\section{Algorithmic solutions}
\label{sec:solutions}

Equation~\ref{eq:Cond} reminds of a 0-1 Knapsack problem, but uses multiplication (rather than summation) for aggregation of items (i.e., security controls) and has a complex utility function (multiplications and summations). This utility function is not order-preserving (i.e., if we add the same control to one set that was riskier than another one, the resulting set may become less risky). Moreover, we see that the budget limit is also a parameter of the utility function itself. This fact complicates the search for the solution.

In other words, although we see some obvious similarities between our problem and the 0-1 knapsack problem family, we need some modifications to the available solutions to find the optimal budget distribution and select cost-efficient security controls.

First, we see our problem as the 0-1 multi-objective knapsack problem \cite{bazgan}, i.e., a 0-1 knapsack problem with many utilities to maximise (i.e., threats to reduce, in our case). 
%For this problem, there are several optimization solutions \cite{bazgan,LOREN}. 
In our previous work \cite{Gabi}, we adapted a dynamic programming solution into our problem. Although this solution was able to solve the problem, in theory, it is not very time- and resource-efficient. Therefore, in this paper we
\begin{itemize}
	\item provide an improved version of the algorithm (e.g., embedding the projection idea), 
	\item adapt a couple of approximate solutions (e.g., greedy and genetic algorithms) to our problem,
	\item conduct an analysis of the applicability of the three solutions.
\end{itemize}

\subsection{Dynamic Programming}

We start with the dynamic programming for solving 0-1 multi-objective knapsack problem proposed by Bazgan et al., \cite{bazgan}. In general, dynamic programming could be applied if the main problem can be decomposed as recursively nested sub-problems. 

First, we enumerate all the elements of $ K $ as  $j=0,1,...,n_{K}$ (where $n_{K}$ is the size of $K$). We will sequentially try security controls, deciding whether to add the latest one into the selected set or reject it. In short, once we reach a control $k_q$, we will have all controls $j=0,...,q$ checked and continue with $j=q+1,...,n_{k}$. 

Costs of controls can be represented as positive integer values, such that $\forall k\in K~(c(j)=C\cdot m_{j})$, where $C$ is the greatest common divisor for all costs and $m_{j}$ is just some positive natural value ($\forall j, m_{j}\in \mathbb{N^+}$). Similarly to controls, we will check the solutions for our problem gradually increasing the limit by $C$ (i.e., $x=C\cdot m$, where $m=0,1,...,m_{max}$).  

To advance in two directions (i.e. considering controls and budget limit) we need an auxiliary matrix $T$. Every cell $T[j][m]$ contains a solution of a sub-problem considering the first $j$ controls and budget limit $x=m\cdot C$ (the overall survival probability $\vec{p}(K_{s}|x),~x=m\cdot C$ computed with Equation~\ref{eq:OverallProb}). Our goal is to consider all security controls and find the optimal budget $x^*=C\cdot m^{*}$ which leads to the minimal value of the first part in Equation~\ref{eq:Cond}. In other words, we are looking for the value (and associated selection of controls) of the cell $T[n_t][m^{*}]$. This will be our solution for the main problem, while every $T[j][m]$ (for $j<n_t$ and $m<m^{*}$) are the sub-problems. 

Since the different combination of controls could fit into the budget limit, there could be several alternatives contributing to one cell in T. In the traditional method for solving 0-1 knapsack problem, every sub-problem (i.e., a selection of alternatives for $T[j][m]$ $j<n_t$ and $m<m^{*}$) could be solved, because the utility function used is order-preserving. Working with a multi-objective optimisation problem, we have no definitive criteria to select the best solution for a sub-problem.

Nevertheless, we may find some criteria which could help us to identify the solutions which are definitely worse than others, to minimise the number of alternatives. We call these solutions as \emph{dominated}, and those solutions for which such a decision cannot be made are called \emph{non-dominated}. Naturally, we should remove all dominated vectors to simplify the computation.

Basically, the core of our algorithm for 0-1 multi-objective knapsack problem could be seen as the following recursive algorithm: 
\begin{figure}[H] %!t
	\includegraphics[width=4.6in]{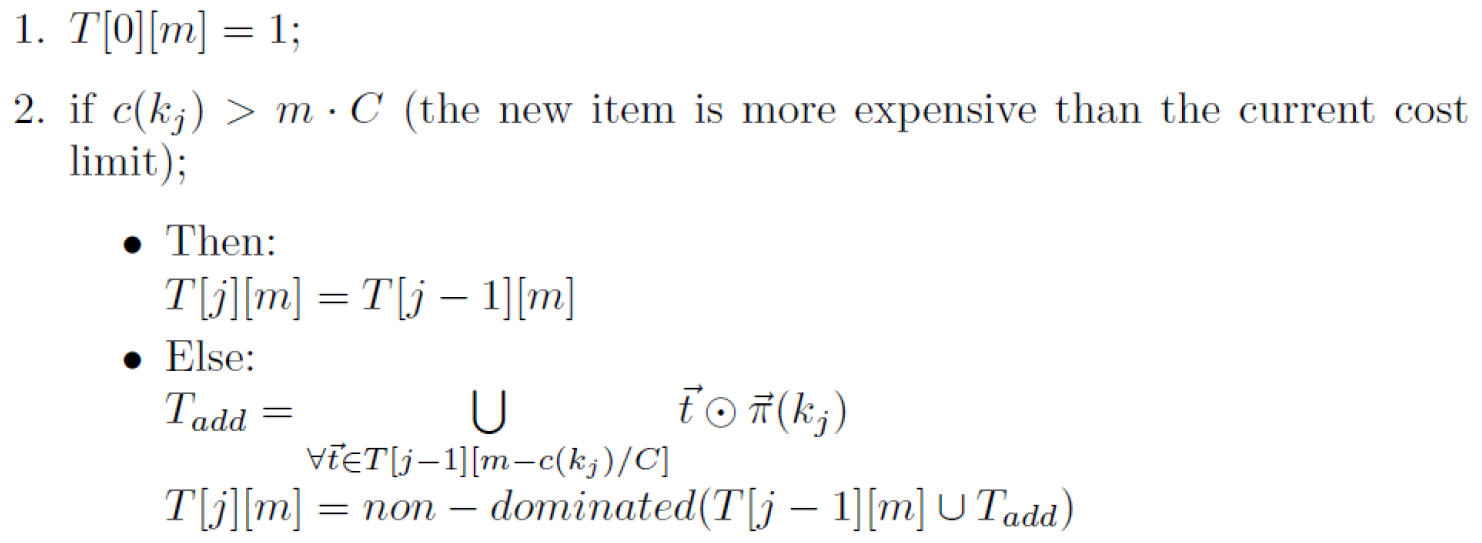}
	\caption{Recursive algorithm}
	\label{fig:Recursive}
\end{figure}
In our problem, we consider that security budget limit is not a given value, but is also a value to be optimised ($x^{*}$) by solving Equation~\ref{eq:Cond}.

It is worth noting that the recursive algorithm does not require the presence of the security investment bound. This fact allows us to start the investment from $ 0 $ and rise it until we find our solution (also extending matrix $T$ for new $x$ to check). In this regard, we should aim \emph{to minimise the number of required iterations and ensure that the solution to Equation~\ref{eq:MaxThird} will be found}.

We can re-write Equation~\ref{eq:MaxThird} as follows, denoting the optimal premium (or residual risk) if $x$ amount is invested in self-protection as $P^{*}(x)$:
\begin{align}
\min\limits_{\forall x}(P^{*}(x) + x).
\label{eq:MaxForth}
\end{align}
Consider some amount of investments $x_{r}\in [0,W^0]$ to be evaluated at step $r\in [0;W^0/C]$. We are interested only in the following future steps y:
\begin{align}
	&x_{r} +P^{*}(x_{r})>x_{r+y} +P^{*}(x_{r+y})\label{eq:FinalCondition};\\
	&x_{r+y}<P^{*}(x_{r})+x_{r} - P^{*}_{min}\label{eq:FinalLimit}, \\
	&where~P^{*}_{min}=\vec{F}\odot \left[ \prod_{\forall k\in K} \vec{\pi}(k)\right])\cdot \vec{L}.
\end{align}
The aforementioned two equations (Equations~\ref{eq:FinalCondition} and \ref{eq:FinalLimit}) enable us to have the following observations. First, Equation~\ref{eq:FinalCondition}, shows how to select the optimal value by comparing the current best value (i.e., up to step $r$) with the next ones ($y>0$). The latter, Equation~\ref{eq:FinalLimit}, defines the stopping point for the algorithm since there will be no more efficient solution if the condition fails. We also may find the first limit, which is: $x^{limit}_{0}=P^{*}(0)-P^{*}_{min}$, considering that $P^{*}_{min}$ is the minimal possible premium/risk that is computed with all possible controls $K_{s}=K$ installed. Naturally, if the company sets a limit for its investments $x^{lim}$ and $P^{*}(x_{r})+x_{r} - P^{*}_{min}>x^{lim}$, we should bound our further steps with $x_{r+y}<x^{lim}$. Note that in this case we are not going to use all invested money, but look for a cost-effective solution within this budget.

We further note that the limit is reset for each better $ x $, since it will be less than the previous one. This observation can be easily proved as follows. Let $x_{r}$ be the previous best value (i.e., for all $r+y-1$ steps) and $x_{r+y}$ be even better than $x_{r}$, i.e.,:
\begin{align}
	P^{*}(x_{r}) + x_{r}>P^{*}(x_{r+y}) + x_{r+y}.
	\label{eq:Step1}
\end{align}
The limits defined at steps $r$ and step $r+y$ are $x^{limit}_{r}$ and $x^{limit}_{r+y}$ consequently:
\begin{align}
	P^{*}(x_{r}) + x_{r}- P^{*}_{min}=x^{limit}_{r}~;~~
	P^{*}(x_{r+y}) + x_{r+y}- P^{*}_{min}=x^{limit}_{r+y}.
	\label{eq:Step2}
\end{align}
We conclude that $x^{limit}_{r}>x^{limit}_{r+y}$.

\paragraph{Algorithm}
Now, we may write an algorithm, based on our ideas described above (Algorithm~\ref{alg:main1}), which a) finds the optimal investments in self-protection $x^*$; b) ensures the lowest expenditure ($(\vec{F}\odot \vec{p}(K_{s}^{*}|x^{*}))\cdot\vec{L}+x^{*}$). The algorithm is based on the dynamic programming approach to solve the 0-1 multi-objective knapsack problem \cite{bazgan}. Although the core part of the algorithm has been re-used, we have adapted it for our problem and make it return the optimal investment as an output, instead of taking it as an input. 

\newcommand\mycommfont[1]{\footnotesize\ttfamily\textcolor{blue}{#1}}
\SetCommentSty{mycommfont}
\SetKwInput{KwInput}{Input} 
\SetKwInput{KwRequire}{Require}                % Set the Input
\SetKwInput{KwEnsure}{Ensure}
\SetKwInput{KwCheck}{Check} 
\SetKwInput{KwCall}{Call}
\SetKwInput{KwReturn}{Return}
\SetKwInput{KwCross}{Crossover}
\SetKwInput{KwMutate}{Mutation}
\SetKwInput{KwMerge}{Merge}
\SetKwInput{KwProcedure}{Procedure}
\SetKwFunction{KwFunction}{Function}

\begin{algorithm}
	\DontPrintSemicolon
	\KwFunction{searchForOptimalInvesments}\;
	\KwInput{$K$, $c$, $\pi$, $\vec{F}$, $\vec{L}$, $x_{init}$, $p_{init}$, C}
	\KwRequire{}
	$K$ \tcp*{a set of controls}
	$c:K\mapsto \mathbb{N}$ \tcp*{cost function}
	$\pi:K\mapsto 2^{[0;1]}$ \tcp*{survival probability per threat function}
	$\vec{F}
	$ \tcp*{frequency vector of $\mathbb{R}^{+}$ values}
	$\vec{L}$ \tcp*{single loss expectency vector of $\mathbb{N}^{+}$ values}
	$x_{init}\in \mathbb{N}$ \tcp*{initial investments}
	$\vec{p}_{init}$ \tcp*{initial probability of survival vector of values from $[0;1]$} \label{line:LastDefinition}
	\KwEnsure{min$(\vec{F}\odot \vec{p}(K_{s}|x))\times \vec{L} + x)$ for optimal security investment $ x^* $}
	$ exp := (\vec{F}\odot\vec{p}_{init})\times\vec{L} + x_{init}$
	\tcp*{Initial expenditure as optimal} \label{line:firstA}
	$P^{*}_{min}:=\vec{F}\odot \left[ \prod_{\forall k\in K} \vec{\pi}(k)\right])\times \vec{L}$\;
	$ x^* := x_{init} $\tcp*{Optimal Investment starts with $x_{init}$}
	$\forall j~T[j][0] :=~\{\vec{p}_{init}\}$
	\tcp*{a dynamic matrix of optimal probabilities. Add new (and the first) column $x=x_{init}$, with one vector $\vec{p}_{init}$} \label{line:Tinit} 
	$C := GCD(\cup_{\forall k\in K} c(k))$ \tcp*{the greatest common divisor for costs}
	$ x := C $ \label{line:xinit}  \tcp*{first increase of investments}
	$ m:=1$ \tcp*{investment counter starts with 1}
	$ n_{k} := |K| $ \tcp*{the size of set $K$} \label{line:finInit}
	\While{$x+x_{init}  \le exp-P^{*}_{min} $\label{line:While}} 
	{
		\tcp*{Do while $x$ is below the optimal expenditure}
		$\forall j~T[j][m] :=~\{\vec{p}_{init}\}$ \tcp*{new column is set with vector $\vec{p}_{init}$}
		
		\For{$j := 1 ~\textbf{to}~ n_{k}$\label{line:For}}  
		{
			\tcp*{for all controls}
			\If{($c(k_j)$ $\leq$ $x$ )\label{line:CompareCost}} 
			{
				\tcp*{check the cost limit}
				$ T[j][m] := non-dominant 
				\left\{ 
				\begin{array}{l}
				\bigcup\limits_{\forall l} \vec{\pi}(k_j) \odot T[j-1][m-c(k_j)/C][l]\\
				T[j-1][m] 
				\end{array}
				\right.$\;
				\tcp*{store all non-dominant vectors comparing two sets: with new control and without.} \label{line:Selection}    
			}
			\Else
			{
				$T[j][m] := T[j-1][m]$\;
				\tcp*{continue without adding new control $j$ } \label{line:HighCost}
			}
		}
		\For{$l:= 0 ~\textbf{to}~ |T[n_{k}][m]|$}
		{
			\tcp*{for all vectors stored in $T[n_{k}][m]$}
			\If{$ (\vec{F}\odot T[n_{k}][m][l])\times \vec{L} + x + x_{init}  < exp $}
			{
				\tcp*{reduced the expenditure?} \label{line:CheckCondition}
				$ exp := (\vec{F}\odot T[n_{k}][m][l])\times \vec{L} + x + x_{init} $\;
				\tcp*{Store this expenditure as optimal} \label{line:reSetCondition}
				$x^* := x$ \tcp*{Remember these investments as optimal} \label{line:reSetOptimal}
			}
		}
		$x:= x + C$ \;
		$m:=m+1$
	}
	\KwReturn{$ exp, x^*$}
	\caption{Selecting the best set of controls}
	\label{alg:main1}
\end{algorithm}
At the initial phase, the algorithm requires all variables and functions for the input. One of the advantages of this algorithm is that it allows taking into account already invested funds $x_{init}$, spent for installation of some initial set of controls $K_{init}$ and reducing the probability of attack with $ \vec{p}_{init} $. The initial controls are not used in our further analysis: $K_{init}\cap K = \emptyset$.

The algorithm starts (\ref{line:firstA} and \ref{line:finInit}) with setting up initial values and computing the minimal premium $P_{min}$ at this point. $GCD(\cup_{\forall k\in K} c(k))$ function returns the Greatest Common Divisor\footnote{The algorithm for finding GCD is well known and is not included in the paper.}.  

We gradually increase the limit $x$ (and its counter $m$) until the procedure reaches the limit $exp-P_{min}$, as Equation~\ref{eq:FinalLimit} states (line~\ref{line:While}). For every limit $x$ we consider all controls one by one (line~\ref{line:For}). For every control, we compare: 1) a set of previously selected controls with $k_{j}$ ($\bigcup\limits_{\forall l} \vec{\pi}(k_j) \odot T[j-1][m-c(k_j)/C][l]$), 2) and the best selection of controls without $k_{j}$ ($T[j-1][m]$)(line~\ref{line:Selection}). We leave only non-dominated elements sets of controls.

As we mentioned earlier, our solution is derived from classic dynamic programming yet, instead of summing the values, we multiply them and look forward to ensuring the lowest overreaching to the maximum. Furthermore, when we set an additional investment for the next column, the overall probability of survival should be computed considering the changes. In particular, if the cost of a control $k_j$ ($c(k_j)$) is higher than the additional investments $x$, we keep the previously selected controls as well as the corresponding survival probabilities $T[j-1][m]$ (line~\ref{line:HighCost}).

After considering all controls for current security investment $x$, we check whether the total expenditure computed by Equation~\ref{eq:FinalCondition} is lower than the previously computed one (line~\ref{line:CheckCondition}). If this condition is met by some vector from $T[n_k][m]$, we reset our optimal security investment to the new one (line~\ref{line:reSetOptimal}) and, more importantly, in line~\ref{line:reSetCondition}, we further replace the expenditure with the new one for further computations (according to the condition in Equation~\ref{eq:FinalLimit}).

We find the lowest possible expenditure and the optimal security investment at the last iteration before the condition in line~\ref{line:While} fails. To find the selected controls, a simple backward algorithm should be applied as we have in Algorithm~\ref{alg:ext} (see Appendix~\ref{app:algos}).

\paragraph{Dominance criteria}
The last part of our algorithm is the criteria for dominance. In our previous work \cite{Gabi}, we used Pareto optimality, i.e., we checked that every element of one vector is better or equal (in our case, lower or equal) than the corresponding element from another vector. We studied further possible ways to improve our DP algorithm since the number of non-dominated vectors grows fast and this slows the algorithm, especially as the number of considered controls grows.

We have found two possible improvements, i) Projection and ii) Sorting. Projection aims to strengthen the dominance criteria, by looking to the remaining controls and assuming the worst case for a vector with lower risk. If the vector under these conditions still results in a lower risk than its opponent, we discard the second vector. Sorting security controls by their cost, leaving the costlier ones for later consideration, tends to reduce the number of possible alternatives (and non-dominated vectors) at the end.  

To implement the projection idea, we first compute the best case values for all controls by multiplying values for threats starting from the latest control. We use a table $Best[ ][ ]$ to store these values, which are computed as:
\begin{align}
Best[j][i] = \prod_{q={n_K-1}}^{j}\pi(k_q)[i]
\end{align}
Next, for every set of already considered controls (e.g., by $j$-th one) we know the lowest survival probability if all remaining controls will be installed. Now, let us compare two vectors $\vec{p}$ and $\vec{p'}$ after considering only $j$-th control and see that risk for the first one is lower than for the second one, i.e.,
\begin{align}
\vec{F}\odot(\vec{p}-\vec{p'}))\times\vec{L} <0.
\end{align}
Let $\mathbb{I}$ be a set of all indexes for threats ($|\mathbb{I}|=n_t$) and for some $i\in \hat{\mathbb{I}} \subset \mathbb{I}$ $p[i]>p'[i]$ and for others $i\in \mathbb{I}\backslash \hat{\mathbb{I}}$ $p[i]\leq p'[i]$. We form a vector $\vec{D}$ with $n_t$ values, as follows:
\begin{align}
D[i]=\left\{ 
\begin{array}{cc}
Best[j][i] &if~i\in \hat{\mathbb{I}}\\
1 & ~if~i\in \mathbb{I}\backslash \hat{\mathbb{I}}
\end{array} 
\right.
\end{align}
We make the situation worse for the first vector by reducing the probabilities of survival for the threats from $\hat{\mathbb{I}}$ and check if the first vector still results in lower risk:
\begin{align}
\vec{F}\odot(\vec{D}\odot(\vec{p}-\vec{p'})))\times\vec{L} <0.
\end{align}
If that is the case, we can remove the second vector from the further consideration, since no matter what we apply in the future the risk computed using the first set of controls will always be lower. Note that this definition of dominance supersedes the former Pareto optimal approach. Algorithm~\ref{alg:Proj} encodes this idea.
\medskip
\begin{algorithm}[!t]
	\DontPrintSemicolon
	\KwFunction{newDominanceCheck}
	\tcp*{Checking new dominance using projection idea}
	\KwInput{$ VectorsA, VectorsB, Best[j], \vec{F},\vec{L}, n_t$}
	\KwRequire{}
	$VectorsA, VectorsB$ \tcp*{two sets of vectors to check for dominance}
	$Best[j]$ \tcp*{A vector with corrective values for the current step j}
	$\vec{F}
	$ \tcp*{frequency vector of $\mathbb{R}^{+}$ values}
	$\vec{L}$ \tcp*{single loss expectency vector of $\mathbb{N}^{+}$ values}
	$n_t$ \tcp*{number of threats}
	\KwEnsure{A set of non-dominated vectors}
	\For{$q < |VectorsA|$}
	{
		\For {$l < |VectorsB|$} 
		{
			$ v1 := 0 $ \tcp*{vector q is dominating}
			$ v2 :=  0$ \tcp*{vector l is dominating}
			\For{$k < n_t$}
			{
				\If {$ (VectorsA[q][i] - VectorsB[l][i]) < 0 $} 
				{
					$ v2 := v2 + (VectorsA[q][i] - VectorsB[l][i]) \cdot F[k ] \cdot L[i]$\;
					$ v1 := v1 + (VectorsA[q][i] - VectorsB[l][i]) \cdot Best[j][i] \cdot F[k ] \cdot L[i] $\;
				}
				\Else
				{
					$ v2 := v2 + (VectorsA[q][i] - VectorsB[l][i]) \cdot Best[j][i] \cdot F[k ] \cdot L[i] $\;
					$ v1 := v1 + (VectorsA[q][i] - VectorsB[l][i]) \cdot F[k ] \cdot L[i] $\;
				}
			}    
			\tcp*{new vector is dominating}
			\If {$ v2 > 0 $} 
			{
				remove vector q from $VectorsA$\;
			}
			\tcp*{old vector is dominating or the same}
			\If {$ v1 <= 0$}
			{
				remove vector l from $VectorsB$\;
			}
		}
	}
	result := $VectorsA \cup VectorsB$
	\caption{Projection Function of the DP}\label{alg:Proj}
\end{algorithm}
\medskip

\subsection{Greedy}
Our DP based solutions accurately find the optimal answer for our main problem, however, they require a lot of time to find the solution when there is a large number of input parameters. Alternatively, approximate solutions could be applied to find nearly optimal answers.

One of such approaches is the Greedy approach \cite{Greedy}. The idea behind the greedy approach is to add (or remove) the elements which separately contribute the most (or the least) to the overall goal. Such an approach does not guarantee to find the optimal value but is fast and easy to implement.

%\paragraph{Algorithm}

Our version of a Greedy algorithm (see Appendix~\ref{app:algos}) starts with all controls to be selected and gradually remove the ones which removal reduces the overall expenditure more than the removal of others. We do this until we are not able to remove any control without increasing the expenditure. 

\subsection{Genetic Algorithm}
One of the most used optimisation approaches is the Genetic Algorithm (GA) \cite{Maya,LOREN,DEWRI,leechoi,VALENTINA}, which provides the optimal or nearest optimal solutions in a short time. We basically keep the foundation of GA \cite{Maya,LOREN}, but make some changes to fit it into our problem (see the Algorithm~\ref{alg:MainFunc} in Appendix~\ref{app:algos}).  

First, we randomly generate an initial population of chromosomes (every chromosome is an example of a security configuration with every gene representing a security control). This population is further used for generation of the new population which will be compared with the initial one with respect to the main criteria (i.e., Equation \ref{eq:MaxThird}).

\begin{figure}[H] %!t
	\includegraphics[width=4.6in]{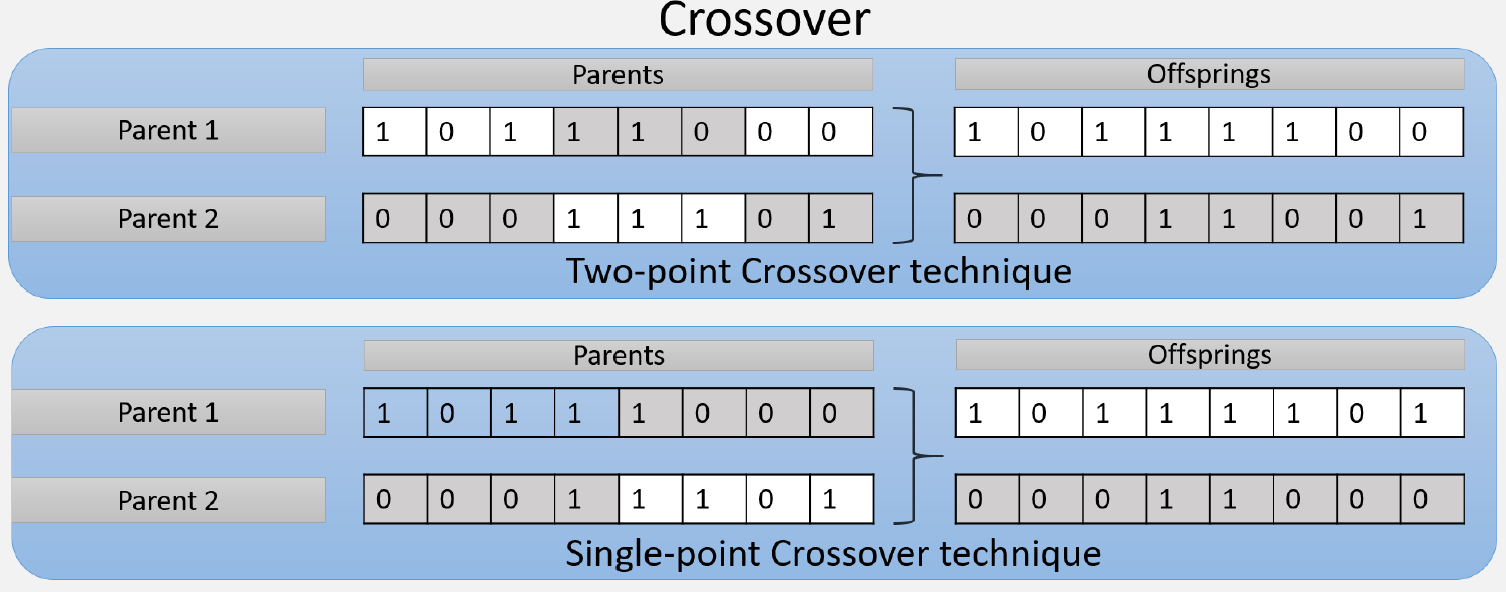}
	\caption{Crossover techniques}
	\label{fig:Cross1}
\end{figure}

The most important part of GA is the creation of a new population (algorithm~\ref{alg:NewPop}), where it comprises crossover and mutation procedures with a merge function. We have improved the techniques for crossover by adapting both two-point and single-point methods as are shown in Figure~\ref{fig:Cross1}. Then, the rest of the population is added to a new population list. Now, the new list of chromosomes will be used for crossover by using three different combinations -- good with good, good with bad and bad with bad. 

\begin{figure}[H] %!t
	\includegraphics[width=4.6in]{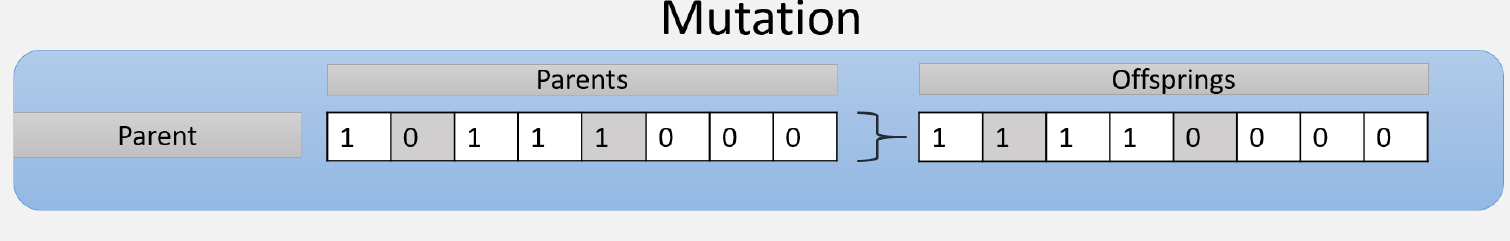}
	\caption{Mutation process}
	\label{fig:Cross2}
\end{figure}

After that, the algorithm sorts the chromosomes and checks if the best chromosome found after the crossover procedure is greater than the one found before. In GA, mutation helps to overcome local minimum solutions and find global optima. We use the basic technique of mutation, selecting random bits (number of random bits is defined in prior) based on a defined number, with elitism technique, and reverse those bits as is shown in the Figure~\ref{fig:Cross2}. This process should be done for the offsprings after crossover procedure, and sort them by their weights to find the best one. 

%% file: 05_CaseStudy.tex
\section{Experiments}
\label{sec:CaseStudy}
In this section, we compare our solution with Greedy and GA in different scenarios. 

The first experiment is to ensure whether the approximate algorithmic solutions and the improved DP yield the same result that we had in our previous work \cite{Gabi}.

Furthermore, what we are interested in is the \emph{execution time} of the solutions and its dependence on various input parameters. The second parameter which we would like to investigate is \emph{accuracy} of the approximate solutions (Greedy and GA). We may expect the approximate solutions to perform better than our DP algorithms, but we would like to test the limits of the algorithms and verify that the version of the DP algorithms with projection idea is indeed beneficial.

To evaluate the accuracy of the approximate solutions, we need to know the exact answer, and therefore we compare the results of the approximate solutions with the result of the DP algorithm. If the computation becomes too slow for the DP-based algorithms to produce the result, we run the GA algorithm several times (e.g., high settings) and assume that the result which we receive most often is the correct one. The last assumption does not provide a 100 percent guarantee that the found result is the exact answer, but this is the best reasonable check we can do. Moreover, for this reason, we do not experiment with the values of the input parameters showing a high dispersion of results.

For the sake of testing, we have created a simple supporting program which randomly generates inputs for our problem. This generation is not fully random, but controlled, i.e., it depends on some quantitative and qualitative input parameters. 

For more complicated cases, we start investigating how the quantity of the input parameters, i.e., the \emph{number of considered controls} and \emph{threats}, affects the execution time and accuracy of the considered solution. Then, we move to the analysis of the effect of quality of input, i.e., how the input is formed. 

We consider the following qualitative input parameters. The \emph{Greatest Common Divisor (GCD)} $C$ determines the granularity of the search for a suitable solution (the higher the GCD is, the easier should be for the algorithms to compare the control costs). \emph{Range of the control costs}, i.e., how different the costs are from each other, could make the solution look too much alike for the algorithms to find a global optimum. The \emph{number of threats affected by a control} shows how many threats could be reduced by installing one control and thus, potentially, could have an impact on the projection idea.   

All experiments have been run on a machine (Intel(R) Core(TM) i7-8550U CPU @ 1.80GHz (8 CPUs), ~2.0GHz) with Windows 10 operating system. We acknowledge that usage of another (e.g., more powerful) computing system will result in different speed of the proposed solutions, but we are more interested in studying the existing dependencies, than finding the exact time of execution for the algorithms. 

\subsection{Basic and Simple scenario}
We use a simple example to demonstrate the work of the proposed solutions. We consider an organisation which has identified five main threats and the corresponding single loss expectancy ($ \vec{L} = \langle 3000, 1800, 2800, 4000, 3800 \rangle $). Eight available controls have been proposed ($|K|=n_{k}=8$) to install with costs as ($c(k_{1}) = 480;~c(k_{2}) = 240;~c(k_{3}) = 120;~c(k_{4}) = 80;~c(k_{5}) = 200;~c(k_{6}) = 120;~c(k_{7}) = 280;~c(k_{8}) = 200$). Finally, the initial probability of survival $\vec{p}_{init}$ (caused by already installed controls), expected frequency of threats $\vec{F}$ and the probabilities of survival for every proposed controls $\vec{\pi}_{j}$ are found as shown in Table~\ref{tab:input}.
\begin{table*}[!ht]
	\begin{center}    
		\resizebox{0.6\textwidth}{!}{
			\begin{tabular}{| c | c | c | c | c | c | c | c | c | c | c | c | c | c | c | c |  c | c | c | c |}
				\hline
				$ \vec{p}_{init} $ &    $ \vec{F} $  &    $ \vec{\pi}_{k1} $ & $ \vec{\pi}_{k2} $ &$ \vec{\pi}_{k3} $&$ \vec{\pi}_{k4} $&$ \vec{\pi}_{k5} $&$ \vec{\pi}_{k6} $&$ \vec{\pi}_{k7} $&$ \vec{\pi}_{k8} $ \\
				\hline
				0.6    &0.8&0.3&    0.9&    0.5&    0.8&    0.9&    0.8&    0.8&    0.6\\
				\hline
				0.7    &0.5 &    0.2    &0.8&    0.7    &    0.6    &    0.5    &0.7    &    0.1    &    0.7\\
				\hline
				0.8    &0.4&    0.5    &    0.9    &    0.9    &    0.9    &    0.8    &0.5    &    0.4    &    0.5\\
				\hline
				0.6&    0.7    &    0.7    &    0.2    &    0.8    &    0.8    &    0.6    &    0.8    &    0.9    &    0.8\\
				\hline
				0.6&    0.5&        0.3    &    0.7    &    0.6    &    0.2    &    0.5    &    0.6    &    0.8    &    0.5\\
				\hline
			\end{tabular}
		}
	\end{center}
	\caption{Input vectors}
	\label{tab:input}
\end{table*}

\begin{figure}[H] %!t
	\includegraphics[width=4.6in, height = 2.2in]{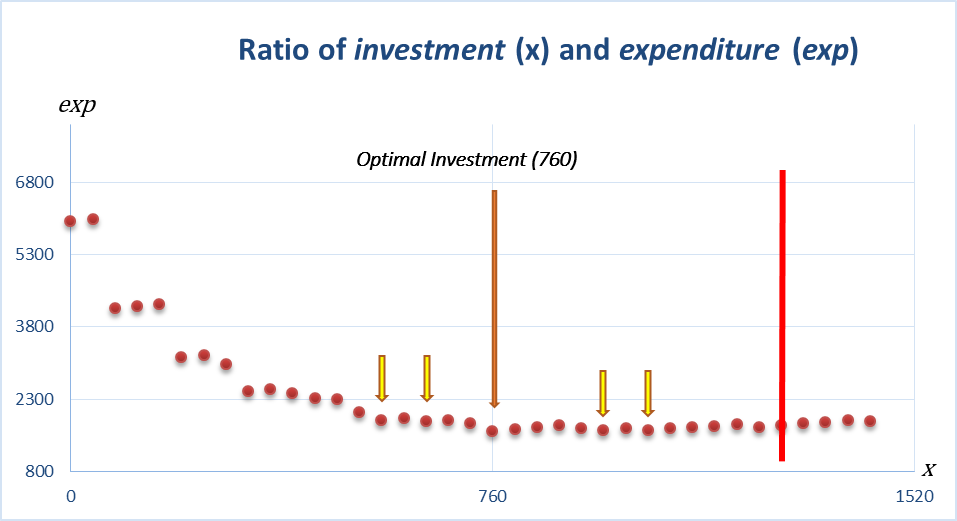}
	\caption{($Exp$) expenditure for security self-investments $x$}
	\label{fig:res}
\end{figure}

As a result of applying all approaches, we obtain the best controls to keep the security expenditure at minimum. Our DP algorithm (see Figure~\ref{fig:res}) successfully passes the local minimums (i.e. for $x=80$, $x=320$ or $x=560$) for expenditure and finds the global one (i.e., for $x=760$). We refer the readers to see equations \ref{eq:Step1} and \ref{eq:Step2} for ensuring how the algorithm finds the global minimum and stops after some steps.
If an organisation selects $\{k_2,k_3,k_4,k_6,k_8\}$ and invests 760, the expenditure is 1642 which is the global minimum so far. Figure \ref{fig:res} also shows that the overall expenditure slowly increases after the optimal value (with some occasional small drops), which confirms that further investments in self-protection are not cost-efficient (even though they reduce the overall risk). Moreover, in order to avoid spending unnecessary resources for the computation, DP based algorithms stop running at $x=1280$.

Both Greedy and GA algorithms find the same optimal answer in a shorter time than DP (as expected). We should underline that the inputs for this toy example are simple, but in the following experiments, we will consider larger and more complex examples in which these algorithms will have difficulty to find the right answer.

\subsection{Quantitative input parameters}
To conduct more complicated experiments, we start with the analysis of the effect of the number of threats $n_t$. The inputs are generated randomly with 20 security controls, 40 GCD, the cost ranging from 80 up to 400, and each control affecting every threat.
Since for GA various settings are possible, in order to limit the number of experiments and focus on the main outcomes, we set ``high\footnote{``High Settings'' for GA represents a larger number of population size and a round of iteration (limit)}'' values for the GA settings (see Table~\ref{tab:Conf1}) during this set of experiments. We will vary some of these settings investigating qualitative input parameters.
\begin{table}[H]
	\resizebox{1\textwidth}{!}{
		\setlength{\tabcolsep}{0.1em}
		\centering
		\begin{tabular}{|c|c|c|c|c|c|c|c|}
			\hline
			\multicolumn{5}{|c|}{}                                                                                                                                                                                                      & \multicolumn{3}{c|}{chromosomes combination} \\ \hline
			Mutation & Limit & \begin{tabular}[c]{@{}c@{}}Population\\ size\end{tabular} & \begin{tabular}[c]{@{}c@{}}Two point \\ crossover \\ percentage\end{tabular} & \begin{tabular}[c]{@{}c@{}}Elitism \\ percentage\end{tabular} & Good \& Good   & Good \& Bad   & Bad \& Bad  \\ \hline
			1 bit        & 1000  & 1000                                                       & 80\%                                                                           & 15\%                                                            & 50\%             & 45\%            & 5\%           \\ \hline
		\end{tabular}
	}
	\caption{Constants for GA execution}
	\label{tab:Conf1}
\end{table}

Table~\ref{tab:data1} shows the results of our experiments. 

\begin{table}[H]
	\caption{Both Time\&Accuracy of the solutions for increasing number of threats}
	\resizebox{1\textwidth}{!}{
		\setlength{\tabcolsep}{0.1em}
		\centering
		\begin{tabular}{ll|l|c|c|c|c|c|c|c|c|c|}
			\cline{3-12}
			&                                      & \textit{Number of threats} & \textbf{5}     & \textbf{10} & \textbf{15}     & \textbf{20} & \textbf{25} & \textbf{30}     & \textbf{35}     & \textbf{40} & \textbf{50} \\ \hline
			\multicolumn{1}{|l|}{\multirow{8}{*}{\textbf{\begin{tabular}[c]{@{}l@{}}20 controls \\ and up to \\ 50 threats\end{tabular}}}} & \multirow{2}{*}{\textbf{DP}}         & {\cellcolor[HTML]{EFEFEF}\textit{Execution time (sec)}}    & {\cellcolor[HTML]{EFEFEF}0.3s}            & {\cellcolor[HTML]{EFEFEF}1.6s}         & {\cellcolor[HTML]{EFEFEF}16.7s}            & {\cellcolor[HTML]{EFEFEF}80.5s}        & {\cellcolor[HTML]{EFEFEF}323s}         & {\cellcolor[HTML]{EFEFEF}756.5s}           &                 &             &             \\ \cline{3-12} 
			\multicolumn{1}{|l|}{}                                                                                                         &                                      & \textit{Overall loss}      & 800.2          & 825.4       & 963.6           & 1140.5      & 1213.8      & 1306.5          &                 &             &             \\ \cline{2-12} 
			\multicolumn{1}{|l|}{}                                                                                                         & \multirow{2}{*}{\textbf{Projection}} & {\cellcolor[HTML]{EFEFEF}\textit{Execution time (sec)}}    & {\cellcolor[HTML]{EFEFEF}0.1s}            & {\cellcolor[HTML]{EFEFEF}0.2s}         & {\cellcolor[HTML]{EFEFEF}0.7s}             & {\cellcolor[HTML]{EFEFEF}6.8s}         & {\cellcolor[HTML]{EFEFEF}8.9s}         & {\cellcolor[HTML]{EFEFEF}18.1s}            & {\cellcolor[HTML]{EFEFEF}27.4s}            & {\cellcolor[HTML]{EFEFEF}58.5s}        & {\cellcolor[HTML]{EFEFEF}271.9s}       \\ \cline{3-12} 
			\multicolumn{1}{|l|}{}                                                                                                         &                                      & \textit{Overall loss}      & 800.2          & 825.4       & 963.6           & 1140.5      & 1213.8      & 1306.5          & 1479            & 1617.8      & 1638.9      \\ \cline{2-12} 
			\multicolumn{1}{|l|}{}                                                                                                         & \multirow{2}{*}{\textbf{Greedy}}     & {\cellcolor[HTML]{EFEFEF}\textit{Execution time (sec)}}    & {\cellcolor[HTML]{EFEFEF}0.01s}           & {\cellcolor[HTML]{EFEFEF}0.014s}       & {\cellcolor[HTML]{EFEFEF}0.021s}           & {\cellcolor[HTML]{EFEFEF}0.015s}       & {\cellcolor[HTML]{EFEFEF}0.008s}       & {\cellcolor[HTML]{EFEFEF}0.009s}           & {\cellcolor[HTML]{EFEFEF}0.0156s}          & {\cellcolor[HTML]{EFEFEF}0.016s}       & {\cellcolor[HTML]{EFEFEF}0.015s}       \\ \cline{3-12} 
			\multicolumn{1}{|l|}{}                                                                                                         &                                      & \textit{Overall loss}      & \textbf{914.3} & 825.4       & \textbf{1045.2} & 1140.5      & 1213.8      & \textbf{1317.1} & \textbf{1560.1} & 1617.8      & 1638.9      \\ \cline{2-12} 
			\multicolumn{1}{|l|}{}                                                                                                         & \multirow{2}{*}{\textbf{GA}}         & {\cellcolor[HTML]{EFEFEF}\textit{Execution time (sec)}}    & {\cellcolor[HTML]{EFEFEF}2.8s}            & {\cellcolor[HTML]{EFEFEF}3.7s}         & {\cellcolor[HTML]{EFEFEF}5.1s}             & {\cellcolor[HTML]{EFEFEF}6.9s}         & {\cellcolor[HTML]{EFEFEF}8.5s}         & {\cellcolor[HTML]{EFEFEF}9.7s}             & {\cellcolor[HTML]{EFEFEF}10.9s}            & {\cellcolor[HTML]{EFEFEF}14.1s}        & {\cellcolor[HTML]{EFEFEF}16.4s}        \\ \cline{3-12} 
			\multicolumn{1}{|l|}{}                                                                                                         &                                      & \textit{Overall loss}      & 800.2          & 825.4       & 963.6           & 1140.5      & 1213.8      & 1306.5          & 1479            & 1617.8      & 1638.9      \\ \hline
		\end{tabular}
	}
	\label{tab:data1}
\end{table}

\begin{figure}[H] %!t
	\includegraphics[width=4.7in, height = 2in]{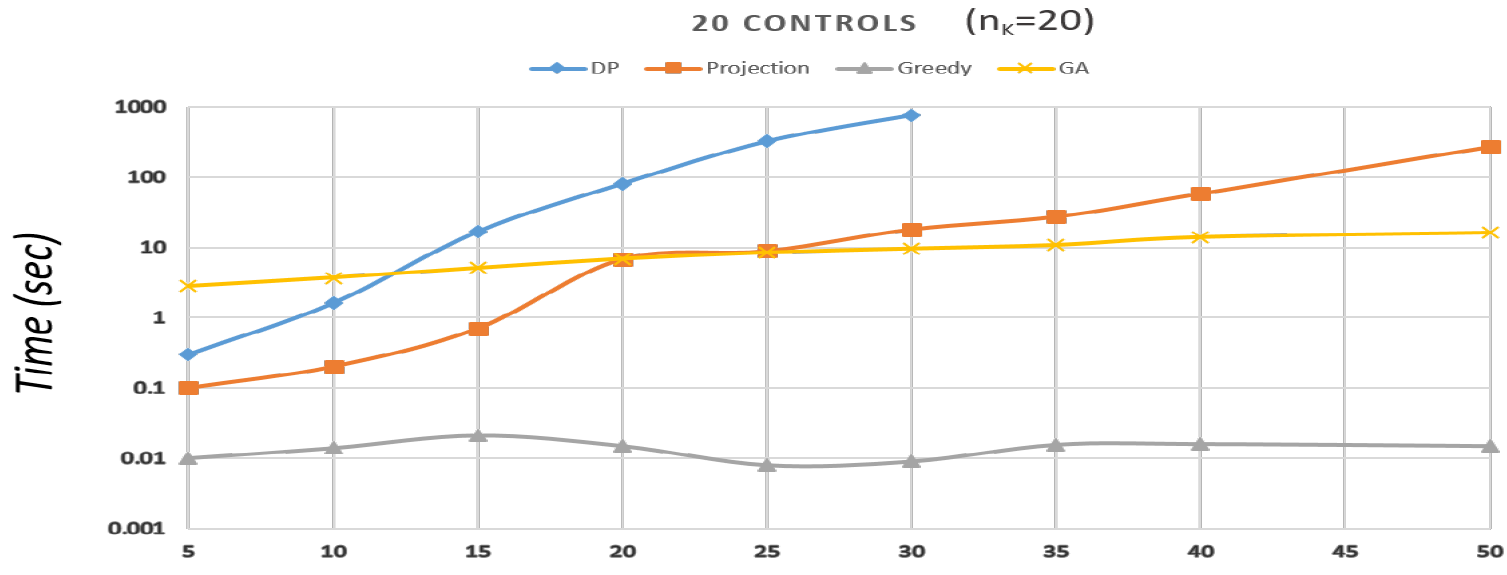}
	\caption{Comparison of execution time for 4 solutions in case of increasing number of threats with 20 control cases}
	\label{fig:Time-Threats}
\end{figure}

First, we analyse the dependency of the execution time on the number of considered threats $n_t$ (see Figure~\ref{fig:Time-Threats}). We see that the Greedy algorithm is not affected very much by the number of threats, the time for execution for others grows with the increase of the number of threats. DP is affected the most and its time of execution rises quickly. Our improved DP algorithm with projection performs better, but still, its time of execution grows faster than the one for GA. 

Although the Greedy solution is the fastest among the four solutions and at the same time it fails more often (highlighted as \emph{\textbf{bold}} in see Table~\ref{tab:Conf1}) to find the optimal answer. GA with our settings performs well and always finds the optimal answer.

\begin{figure}[H] %!t
	\includegraphics[width=4.9in, height = 3.5in]{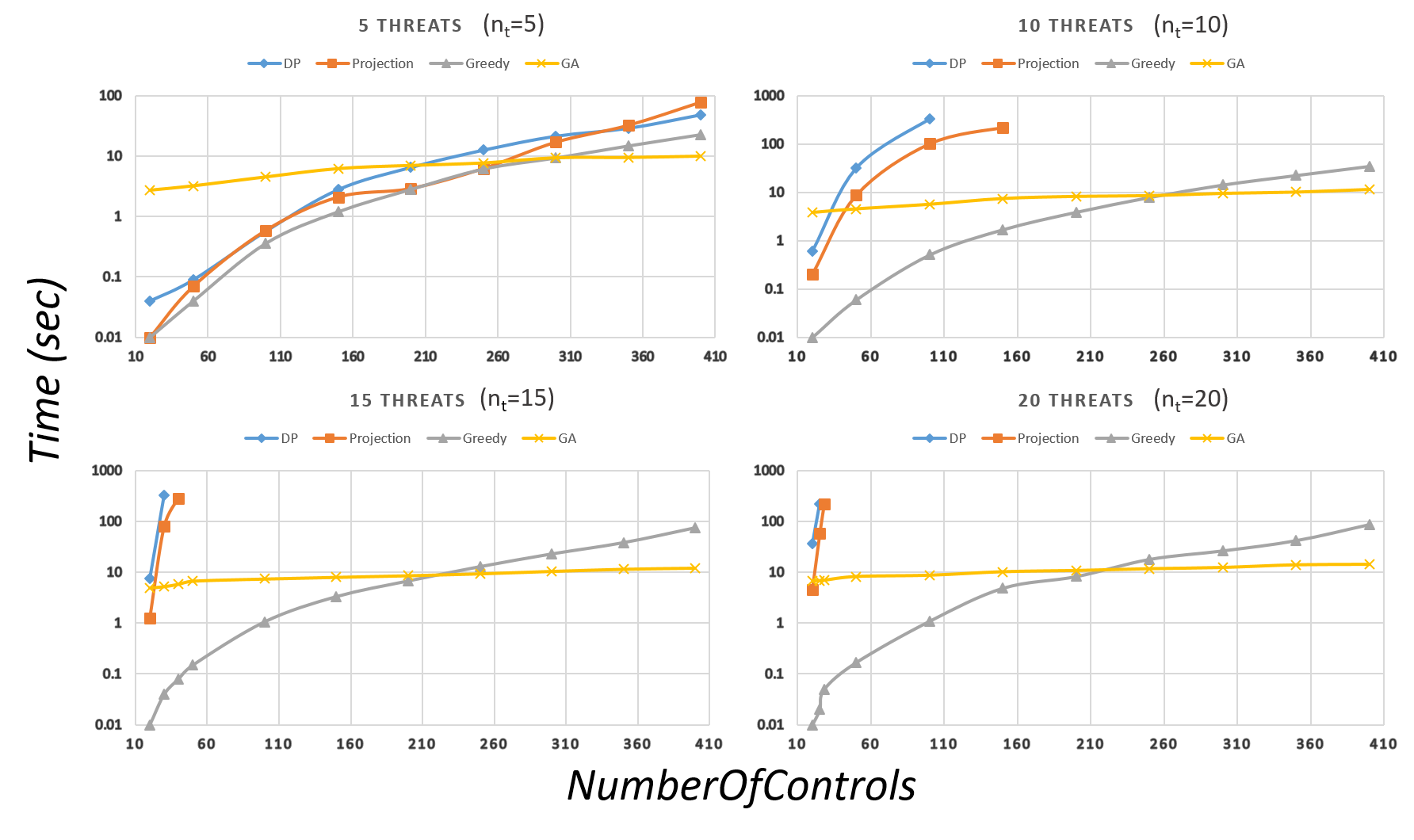}
	\caption{Comparison of execution time for 4 solutions in case of increasing number of controls with 5, 10, 15, 20 threats cases}
	\label{fig:comparL}
\end{figure}

Next, we conduct 4 series of experiments, to analyse the dependency of the execution time on the number of controls. The number of threats considered is fixed in every experiment $n_t=5,10,15$ and $20$. The four graphs in 
Figure~\ref{fig:comparL} represent the results of the time dependency while the Table~\ref{tab:data} indicates both time and accuracy of the solutions.
\begin{table}[t]
	\caption{Both Time\&Accuracy comparison for increasing number of controls}
	\resizebox{1\textwidth}{!}{
		\setlength{\tabcolsep}{0.1em}
		\centering
		\begin{tabular}{|c|c|c|c|c|c|c|c|c|c|c|c|}
			\hline
			&                                      & \textit{Number of Controls} & \textbf{20} & \textbf{50}    & \textbf{100}   & \textbf{150}   & \textbf{200}   & \textbf{250}   & \textbf{300}   & \textbf{350}   & \textbf{400}   \\ \hline
			\multirow{8}{*}{\textbf{\begin{tabular}[c]{@{}c@{}}5 threats case with \\ increasing number \\ of controls\end{tabular}}}  & \multirow{2}{*}{\textit{DP}}         & {\cellcolor[HTML]{EFEFEF}\textit{Execution time (sec)}}     & {\cellcolor[HTML]{EFEFEF}0.04s}        & {\cellcolor[HTML]{EFEFEF}0.09s}           & {\cellcolor[HTML]{EFEFEF}0.57s}           & {\cellcolor[HTML]{EFEFEF}2.8s}            & {\cellcolor[HTML]{EFEFEF}6.5s}            & {\cellcolor[HTML]{EFEFEF}12.6s}           & {\cellcolor[HTML]{EFEFEF}21.4s}           & {\cellcolor[HTML]{EFEFEF}28.9s}           & {\cellcolor[HTML]{EFEFEF}48.3s}           \\ \cline{3-12} 
			&                                      & \textit{Overall loss}       & 510.6       & 359.6          & 356.9          & 329.3          & 322.3          & 322.3          & 322.3          & 322.3          & 322.3          \\ \cline{2-12} 
			& \multirow{2}{*}{\textit{Projection}} & {\cellcolor[HTML]{EFEFEF}\textit{Execution time (sec)}}     & {\cellcolor[HTML]{EFEFEF}0.01s}        & {\cellcolor[HTML]{EFEFEF}0.07s}           & {\cellcolor[HTML]{EFEFEF}0.59s}           & {\cellcolor[HTML]{EFEFEF}2.1s}            & {\cellcolor[HTML]{EFEFEF}2.9s}            & {\cellcolor[HTML]{EFEFEF}6.2s}            & {\cellcolor[HTML]{EFEFEF}16.9s}           & {\cellcolor[HTML]{EFEFEF}32.1s}           & {\cellcolor[HTML]{EFEFEF}77.4s}           \\ \cline{3-12} 
			&                                      & \textit{Overall loss}       & 510.6       & 359.6          & 356.9          & 329.3          & 322.3          & 322.3          & 322.3          & 322.3          & 322.3          \\ \cline{2-12} 
			& \multirow{2}{*}{\textit{Greedy}}     & {\cellcolor[HTML]{EFEFEF}\textit{Execution time (sec)}}     & {\cellcolor[HTML]{EFEFEF}0.01s}        & {\cellcolor[HTML]{EFEFEF}0.04s}           & {\cellcolor[HTML]{EFEFEF}0.36s}           & {\cellcolor[HTML]{EFEFEF}1.2s}            & {\cellcolor[HTML]{EFEFEF}2.8s}            & {\cellcolor[HTML]{EFEFEF}6.1s}            & {\cellcolor[HTML]{EFEFEF}9.3s}            & {\cellcolor[HTML]{EFEFEF}14.7s}           & {\cellcolor[HTML]{EFEFEF}22.6s}           \\ \cline{3-12} 
			&                                      & \textit{Overall loss}       & 510.6       & \textbf{413}   & 356.9          & \textbf{353.7} & \textbf{337.9} & \textbf{337.9} & \textbf{337.9} & \textbf{342.8} & \textbf{363.7} \\ \cline{2-12} 
			& \multirow{2}{*}{\textit{GA}}         & {\cellcolor[HTML]{EFEFEF}\textit{Execution time (sec)}}     & {\cellcolor[HTML]{EFEFEF}2.7s}         & {\cellcolor[HTML]{EFEFEF}3.2s}            & {\cellcolor[HTML]{EFEFEF}4.5s}            & {\cellcolor[HTML]{EFEFEF}6.2s}            & {\cellcolor[HTML]{EFEFEF}7s}              & {\cellcolor[HTML]{EFEFEF}7.69s}           & {\cellcolor[HTML]{EFEFEF}9.4s}            & {\cellcolor[HTML]{EFEFEF}9.5s}            & {\cellcolor[HTML]{EFEFEF}10s}             \\ \cline{3-12} 
			&                                      & \textit{Overall loss}       & 510.6       & 359.6          & 356.9          & 329.3          & 322.3          & 322.3          & 322.3          & 322.3          & 322.3          \\ \hline
			\multirow{8}{*}{\textbf{\begin{tabular}[c]{@{}c@{}}10 threats case with \\ increasing number \\ of controls\end{tabular}}} & \multirow{2}{*}{\textit{DP}}         & {\cellcolor[HTML]{EFEFEF}\textit{Execution time (sec)}}     & {\cellcolor[HTML]{EFEFEF}0.6s}         & {\cellcolor[HTML]{EFEFEF}31.9s}           & {\cellcolor[HTML]{EFEFEF}331.4s}          &                &                &                &                &                &                \\ \cline{3-12} 
			&                                      & \textit{Overall loss}       & 850.1       & 553.7          & 468.1          &                &                &                &                &                &                \\ \cline{2-12} 
			& \multirow{2}{*}{\textit{Projection}} & {\cellcolor[HTML]{EFEFEF}\textit{Execution time (sec)}}     & {\cellcolor[HTML]{EFEFEF}0.2s}         & {\cellcolor[HTML]{EFEFEF}8.7s}            & {\cellcolor[HTML]{EFEFEF}101s}            & {\cellcolor[HTML]{EFEFEF}220.7s}          &                &                &                &                &                \\ \cline{3-12} 
			&                                      & \textit{Overall loss}       & 850.1       & 553.7          & 468.1          & 428.6          &                &                &                &                &                \\ \cline{2-12} 
			& \multirow{2}{*}{\textit{Greedy}}     & {\cellcolor[HTML]{EFEFEF}\textit{Execution time (sec)}}     & {\cellcolor[HTML]{EFEFEF}0.01s}        & {\cellcolor[HTML]{EFEFEF}0.06s}           & {\cellcolor[HTML]{EFEFEF}0.51s}           & {\cellcolor[HTML]{EFEFEF}1.7s}            & {\cellcolor[HTML]{EFEFEF}3.9s}            & {\cellcolor[HTML]{EFEFEF}7.9s}            & {\cellcolor[HTML]{EFEFEF}14.3s}           & {\cellcolor[HTML]{EFEFEF}22.5s}           & {\cellcolor[HTML]{EFEFEF}34.8s}           \\ \cline{3-12} 
			&                                      & \textit{Overall loss}       & 850.1       & \textbf{562.7} & \textbf{488.3} & 428.6          & 416.5          & 385.1          & \textbf{422.3}          & \textbf{422.3}          & \textbf{455.3}          \\ \cline{2-12} 
			& \multirow{2}{*}{\textit{GA}}         & {\cellcolor[HTML]{EFEFEF}\textit{Execution time (sec)}}     & {\cellcolor[HTML]{EFEFEF}3.77s}        & {\cellcolor[HTML]{EFEFEF}4.48s}           & {\cellcolor[HTML]{EFEFEF}5.6s}            & {\cellcolor[HTML]{EFEFEF}7.4s}            & {\cellcolor[HTML]{EFEFEF}8.2s}            & {\cellcolor[HTML]{EFEFEF}8.5s}            & {\cellcolor[HTML]{EFEFEF}9.5s}            & {\cellcolor[HTML]{EFEFEF}10.1s}           & {\cellcolor[HTML]{EFEFEF}11.5s}           \\ \cline{3-12} 
			&                                      & \textit{Overall loss}       & 850.1       & 553.7          & 468.1          & 428.6          & 416.5          & 385.1          & 385.1          & 385.1          & 385.1          \\ \hline
			\multirow{8}{*}{\textbf{\begin{tabular}[c]{@{}c@{}}15 threats case with \\ increasing number \\ of controls\end{tabular}}} & \multirow{2}{*}{\textit{DP}}         & {\cellcolor[HTML]{EFEFEF}\textit{Execution time (sec)}}     & {\cellcolor[HTML]{EFEFEF}7.6s}         &                &                &                &                &                &                &                &                \\ \cline{3-12} 
			&                                      & \textit{Overall loss}       & 977         &                &                &                &                &                &                &                &                \\ \cline{2-12} 
			& \multirow{2}{*}{\textit{Projection}} & {\cellcolor[HTML]{EFEFEF}\textit{Execution time (sec)}}     & {\cellcolor[HTML]{EFEFEF}1.24s}        &                &                &                &                &                &                &                &                \\ \cline{3-12} 
			&                                      & \textit{Overall loss}       & 977         &                &                &                &                &                &                &                &                \\ \cline{2-12} 
			& \multirow{2}{*}{\textit{Greedy}}     & {\cellcolor[HTML]{EFEFEF}\textit{Execution time (sec)}}     & {\cellcolor[HTML]{EFEFEF}0.01s}        & {\cellcolor[HTML]{EFEFEF}0.15s}           & {\cellcolor[HTML]{EFEFEF}1.06s}           & {\cellcolor[HTML]{EFEFEF}3.31s}           & {\cellcolor[HTML]{EFEFEF}6.7s}            & {\cellcolor[HTML]{EFEFEF}12.8s}           & {\cellcolor[HTML]{EFEFEF}22.7s}           & {\cellcolor[HTML]{EFEFEF}37.8s}           & {\cellcolor[HTML]{EFEFEF}74s}             \\ \cline{3-12} 
			&                                      & \textit{Overall loss}       & 977         & \textbf{834.2}          & 598.8          & \textbf{599.2}          & 466.4          & \textbf{478.4}          & \textbf{488.5}          & \textbf{471.9}          & \textbf{441.9}          \\ \cline{2-12} 
			& \multirow{2}{*}{\textit{GA}}         & {\cellcolor[HTML]{EFEFEF}\textit{Execution time (sec)}}     & {\cellcolor[HTML]{EFEFEF}4.9s}         & {\cellcolor[HTML]{EFEFEF}6.7s}            & {\cellcolor[HTML]{EFEFEF}7.4s}            & {\cellcolor[HTML]{EFEFEF}8s}              & {\cellcolor[HTML]{EFEFEF}8.6s}            & {\cellcolor[HTML]{EFEFEF}9.4s}            & {\cellcolor[HTML]{EFEFEF}10.5s}           & {\cellcolor[HTML]{EFEFEF}11.6s}           & {\cellcolor[HTML]{EFEFEF}12.2s}           \\ \cline{3-12} 
			&                                      & \textit{Overall loss}       & 977         & 789.7          & 598.8          & 576.2          & 466.4          & 466.4          & 468.5          & 465.2          & 434            \\ \hline
			\multirow{8}{*}{\textbf{\begin{tabular}[c]{@{}c@{}}20 threats case with \\ increasing number \\ of controls\end{tabular}}} & \multirow{2}{*}{\textit{DP}}         & {\cellcolor[HTML]{EFEFEF}\textit{Execution time (sec)}}     & {\cellcolor[HTML]{EFEFEF}36.7s}        &                &                &                &                &                &                &                &                \\ \cline{3-12} 
			&                                      & \textit{Overall loss}       & 1236.2      &                &                &                &                &                &                &                &                \\ \cline{2-12} 
			& \multirow{2}{*}{\textit{Projection}} & {\cellcolor[HTML]{EFEFEF}\textit{Execution time (sec)}}     & {\cellcolor[HTML]{EFEFEF}4.6s}         &                &                &                &                &                &                &                &                \\ \cline{3-12} 
			&                                      & \textit{Overall loss}       & 1236.2      &                &                &                &                &                &                &                &                \\ \cline{2-12} 
			& \multirow{2}{*}{\textit{Greedy}}     & {\cellcolor[HTML]{EFEFEF}\textit{Execution time (sec)}}     & {\cellcolor[HTML]{EFEFEF}0.01s}        & {\cellcolor[HTML]{EFEFEF}0.17s}           & {\cellcolor[HTML]{EFEFEF}1.1s}            & {\cellcolor[HTML]{EFEFEF}4.9s}            & {\cellcolor[HTML]{EFEFEF}8.3s}            & {\cellcolor[HTML]{EFEFEF}18.1s}           & {\cellcolor[HTML]{EFEFEF}26.6s}           & {\cellcolor[HTML]{EFEFEF}42.3s}           & {\cellcolor[HTML]{EFEFEF}87.5s}           \\ \cline{3-12} 
			&                                      & \textit{Overall loss}       & 1236.2      & 850.9          & \textbf{774.2}          & \textbf{699.2}          & \textbf{678.6}          & \textbf{645.3}          & \textbf{653.8}          & \textbf{667.7}          & 577.9            \\ \cline{2-12} 
			& \multirow{2}{*}{\textit{GA}}         & {\cellcolor[HTML]{EFEFEF}\textit{Execution time (sec)}}     & {\cellcolor[HTML]{EFEFEF}6.7s}         & {\cellcolor[HTML]{EFEFEF}8.2s}            & {\cellcolor[HTML]{EFEFEF}8.7s}            & {\cellcolor[HTML]{EFEFEF}10.2s}           & {\cellcolor[HTML]{EFEFEF}10.8s}           & {\cellcolor[HTML]{EFEFEF}11.7s}           & {\cellcolor[HTML]{EFEFEF}12.4s}           & {\cellcolor[HTML]{EFEFEF}14s}             & {\cellcolor[HTML]{EFEFEF}14.3s}           \\ \cline{3-12} 
			&                                      & \textit{Overall loss}       & 1236.2      & 850.9          & 714.2          & 659.1          & 642            & 619.2          & 619.2          & 612.1          & 577.9          \\ \hline
		\end{tabular}
	}
	\label{tab:data}
\end{table}
The Greedy solution in all cases performs faster than others for a lower number of controls, but then its execution time increases. For instance, its execution time grows faster than GA, which outperforms the Greedy solution as the number of controls rises larger than about 200. Moreover, we see that GA is affected by the number of controls only slightly, and, thus, we may expect that its performance continues to stay the best and the execution time relatively fast.

DP algorithms, as was expected, are the slowest and are the most affected by the number of controls. Their execution time grows as $n_K$ grows especially when the number of considered threats is large. DP with projection outperforms the ordinary DP algorithm, but this difference is much lower than the difference with approximate algorithms.

Analysing accuracy, we can see that the Greedy solution often fails to yield the optimal answer whereas GA always finds the best ones. We admit that, at some points, it is complicated to say whether GA provides the optimal answer, however, we are confident because of the several times of running the GA.

Thus, we may conclude that the projection idea helps to improve the DP algorithm, but its effect is reducing with the increase of the input parameters. Also, Greedy and GA outperform the DP solutions and are able to provide the result in a reasonable time even for a high number of threats and controls. GA algorithm is less affected by the grows of parameters, although the Greedy one is faster for smaller numbers. But, the Greedy algorithm usually provides an only nearly optimal answer.

\subsection{Qualitative parameters}

We further analyse the parameters which measure \emph{how} the input is formed. For every of the three selected qualitative input parameters, we conduct at least 3 experiments with different values of these parameters, keeping the others constant. The results of the experiments can be found in Table~\ref{tab:data2}.
\begin{table}[H]
	\resizebox{1\textwidth}{!}{
		\setlength{\tabcolsep}{0.1em}
		\centering
		\begin{tabular}{cc|c|c|c|c|c|}
			\cline{3-7}
			&                                          & \textit{Algorithms}     & \textbf{DP}       & \textbf{Projection} & \textbf{Greedy} & \textbf{GA} \\ \hline
			\multicolumn{1}{|l|}{\multirow{6}{*}{\textit{\textbf{\begin{tabular}[c]{@{}l@{}}Changing GCD\\ (30 controls and \\ 10 threats with \\ 80 to 160 cost range)\end{tabular}}}}}                   & \multirow{2}{*}{\textbf{40 GCD}}         & {\cellcolor[HTML]{EFEFEF}\textit{Execution time (sec)}} & {\cellcolor[HTML]{EFEFEF}418s}               & {\cellcolor[HTML]{EFEFEF}74.6s}                & {\cellcolor[HTML]{EFEFEF}0.03s}            & {\cellcolor[HTML]{EFEFEF}3.8s}         \\ \cline{3-7} 
			\multicolumn{1}{|l|}{}                                                                                                                                                                         &                                          & \textit{Overall loss}   & 627.1             & 627.1               & \textbf{700.6}  & 627.1       \\ \cline{2-7} 
			\multicolumn{1}{|l|}{}                                                                                                                                                                         & \multirow{2}{*}{\textbf{20 GCD}}         & {\cellcolor[HTML]{EFEFEF}\textit{Execution time (sec)}} & {\cellcolor[HTML]{EFEFEF}496s}               & {\cellcolor[HTML]{EFEFEF}75.5s}                & {\cellcolor[HTML]{EFEFEF}0.03s}            & {\cellcolor[HTML]{EFEFEF}4.2s}         \\ \cline{3-7} 
			\multicolumn{1}{|l|}{}                                                                                                                                                                         &                                          & \textit{Overall loss}   & 543.9             & 543.9               & \textbf{567.8}  & 543.9       \\ \cline{2-7} 
			\multicolumn{1}{|l|}{}                                                                                                                                                                         & \multirow{2}{*}{\textbf{10 GCD}}         & {\cellcolor[HTML]{EFEFEF}\textit{Execution time (sec)}} & {\cellcolor[HTML]{EFEFEF}1369.6s}            & {\cellcolor[HTML]{EFEFEF}92.1s}                & {\cellcolor[HTML]{EFEFEF}0.03s}            & {\cellcolor[HTML]{EFEFEF}4.8s}         \\ \cline{3-7} 
			\multicolumn{1}{|l|}{}                                                                                                                                                                         &                                          & \textit{Overall loss}   & 576.9             & 576.9               & \textbf{613.2}  & 576.9       \\ \hline
			\multicolumn{1}{|l|}{\multirow{6}{*}{\textit{\textbf{\begin{tabular}[c]{@{}l@{}}Different range \\ of cost\\ (30 controls and \\ 10 threats with \\ 40 GCD)\end{tabular}}}}}                   & \multirow{2}{*}{\textbf{80 - 400 range}} & {\cellcolor[HTML]{EFEFEF}\textit{Execution time (sec)}} & {\cellcolor[HTML]{EFEFEF}6s}                 & {\cellcolor[HTML]{EFEFEF}1.2s}                 & {\cellcolor[HTML]{EFEFEF}0.01s}            & {\cellcolor[HTML]{EFEFEF}3.6s}         \\ \cline{3-7} 
			\multicolumn{1}{|l|}{}                                                                                                                                                                         &                                          & \textit{Overall loss}   & 706.6             & 706.6               & 706.6           & 706.6       \\ \cline{2-7} 
			\multicolumn{1}{|l|}{}                                                                                                                                                                         & \multirow{2}{*}{\textbf{80 - 160 range}} & {\cellcolor[HTML]{EFEFEF}\textit{Execution time (sec)}} & {\cellcolor[HTML]{EFEFEF}418s}               & {\cellcolor[HTML]{EFEFEF}74.6s}                & {\cellcolor[HTML]{EFEFEF}0.03s}            & {\cellcolor[HTML]{EFEFEF}3.8s}         \\ \cline{3-7} 
			\multicolumn{1}{|l|}{}                                                                                                                                                                         &                                          & \textit{Overall loss}   & 627.1             & 627.1               & \textbf{700.6}  & 627.1       \\ \cline{2-7} 
			\multicolumn{1}{|l|}{}                                                                                                                                                                         & \multirow{2}{*}{\textbf{80 - 120 range}} & {\cellcolor[HTML]{EFEFEF}\textit{Execution time (sec)}} & {\cellcolor[HTML]{EFEFEF}1741.7s}            & {\cellcolor[HTML]{EFEFEF}132.5s}               & {\cellcolor[HTML]{EFEFEF}0.03s}            & {\cellcolor[HTML]{EFEFEF}4.2s}         \\ \cline{3-7} 
			\multicolumn{1}{|l|}{}                                                                                                                                                                         &                                          & \textit{Overall loss}   & 495.3             & 495.3               & \textbf{544.2}  & 495.3       \\ \hline
			\multicolumn{1}{|l|}{\multirow{8}{*}{\textit{\textbf{\begin{tabular}[c]{@{}l@{}}Affected threats\\ (30 controls and \\ 10 threats with \\ 40 GCD and 80 to \\ 400 cost range)\end{tabular}}}}} & \multirow{2}{*}{\textbf{1 threat}}       & {\cellcolor[HTML]{EFEFEF}\textit{Execution time (sec)}} & {\cellcolor[HTML]{EFEFEF}\textgreater{}600s} & {\cellcolor[HTML]{EFEFEF}0.2s}                 & {\cellcolor[HTML]{EFEFEF}0.01s}            & {\cellcolor[HTML]{EFEFEF}4.7s}         \\ \cline{3-7} 
			\multicolumn{1}{|l|}{}                                                                                                                                                                         &                                          & \textit{Overall loss}   &                   & 10277               & 10277           & 10277       \\ \cline{2-7} 
			\multicolumn{1}{|l|}{}                                                                                                                                                                         & \multirow{2}{*}{\textbf{4 threats}}      & {\cellcolor[HTML]{EFEFEF}\textit{Execution time (sec)}} & {\cellcolor[HTML]{EFEFEF}\textgreater{}600s} & {\cellcolor[HTML]{EFEFEF}3.3s}                 & {\cellcolor[HTML]{EFEFEF}0.01s}            & {\cellcolor[HTML]{EFEFEF}3.8s}         \\ \cline{3-7} 
			\multicolumn{1}{|l|}{}                                                                                                                                                                         &                                          & \textit{Overall loss}   &                   & 2411.5              & 2411.5          & 2411.5      \\ \cline{2-7} 
			\multicolumn{1}{|l|}{}                                                                                                                                                                         & \multirow{2}{*}{\textbf{7 threats}}      & {\cellcolor[HTML]{EFEFEF}\textit{Execution time (sec)}} & {\cellcolor[HTML]{EFEFEF}339.9s}             & {\cellcolor[HTML]{EFEFEF}1.7s}                 & {\cellcolor[HTML]{EFEFEF}0.01s}            & {\cellcolor[HTML]{EFEFEF}3.6s}         \\ \cline{3-7} 
			\multicolumn{1}{|l|}{}                                                                                                                                                                         &                                          & \textit{Overall loss}   & 947.2             & 947.2               & 947.2           & 947.2       \\ \cline{2-7} 
			\multicolumn{1}{|l|}{}                                                                                                                                                                         & \multirow{2}{*}{\textbf{10 threats}}     & {\cellcolor[HTML]{EFEFEF}\textit{Execution time (sec)}} & {\cellcolor[HTML]{EFEFEF}6s}                 & {\cellcolor[HTML]{EFEFEF}1.2s}                 & {\cellcolor[HTML]{EFEFEF}0.01s}            & {\cellcolor[HTML]{EFEFEF}3.6s}         \\ \cline{3-7} 
			\multicolumn{1}{|l|}{}                                                                                                                                                                         &                                          & \textit{Overall loss}   & 706.6             & 706.6               & 706.6           & 706.6       \\ \hline
		\end{tabular}
	}
	\caption{Both time and accuracy of the solutions for different scenarios}
	\label{tab:data2}
\end{table}

We run the experiment with 30 controls and 10 threats, varying GCD (40, 20, and 10) used for the input (cost, in this case) generation. The cost of controls varies between 80 and 160. We see that GA is only slightly affected by the variation of GCD, but DP algorithms slow down quickly with the decrease of this parameter. The greedy algorithm does not change the time of its execution, as it does not depend on GCD. With these settings, we also see that the Greedy algorithm is not able to find the optimal answer.

We also analysed how the execution time depends on the variation of the control costs (see Table~\ref{tab:data2}. Again, we see that DP-based algorithms significantly slow down as the variance decreases (the variant with projection slows down more slowly). GA also takes slightly more time once the variance becomes lower. Also, the Greedy algorithm slows slightly down, but its precision drops with a lower variance in control costs.   

Finally, we let every control affect only a certain number of threats (e.g., 1, 4, 7, and 10). The original DP solution requires significantly more time if fewer threats affected, while GA and DP with projection solutions' time gradually drops. An interesting observation we have got for DP with projection: it finds the solution much faster for 1 threat affected case and immediately increases for 4 threats affected case. This can be concluded that 1 threat affected scenario is much simpler for the projection idea since it has more non-dominating vectors to project. For Greedy, it is almost constant around 0.01 seconds for all cases. In this experiment, we see that the input parameters are good enough for the Greedy algorithm to find the optimal answer. 
\subsection{Precision of GA and Greedy algorithms}
We see that GA and Greedy algorithms are faster than DP ones but they are imprecise. In other words, we cannot be sure that the result they produce is the optimal answer to the problem. Therefore, we would like to harden the input parameters for the approximate solutions to test their limits. 
Moreover, we are going to lower the settings for GA (limit number and population size) to test them.
\begin{table}[H]
	\resizebox{\textwidth}{!}{%
		\begin{tabular}{lclllllllllll}
			\hline
			\multicolumn{12}{c}{\textbf{Genetic Algorithm}} &
			\textbf{Greedy} \\ \hline
			\multicolumn{1}{|l|}{\textbf{}} &
			\multicolumn{1}{c|}{\textbf{}} &
			\multicolumn{1}{l|}{\textit{\textbf{Limit number}}} &
			\multicolumn{3}{c|}{\textbf{100}} &
			\multicolumn{3}{c|}{\textbf{500}} &
			\multicolumn{3}{c|}{\textbf{1000}} &
			\multicolumn{1}{l|}{} \\ \hline
			\multicolumn{1}{|l|}{\textbf{}} &
			\multicolumn{1}{c|}{\textbf{}} &
			\multicolumn{1}{l|}{\textit{\textbf{Population size}}} &
			\multicolumn{1}{c|}{\textit{\textbf{100}}} &
			\multicolumn{1}{c|}{\textit{\textbf{500}}} &
			\multicolumn{1}{c|}{\textit{\textbf{1000}}} &
			\multicolumn{1}{c|}{\textit{\textbf{100}}} &
			\multicolumn{1}{c|}{\textit{\textbf{500}}} &
			\multicolumn{1}{c|}{\textit{\textbf{1000}}} &
			\multicolumn{1}{c|}{\textit{\textbf{100}}} &
			\multicolumn{1}{c|}{\textit{\textbf{500}}} &
			\multicolumn{1}{c|}{\textit{\textbf{1000}}} &
			\multicolumn{1}{l|}{} \\  \hline
			\multicolumn{1}{|l|}{\multirow{6}{*}{\textbf{\begin{tabular}[c]{@{}l@{}}Increasing number \\ of controls\\ (20 threats \\ 40 gcd with range \\ of 80 to 400 costs)\end{tabular}}}} &
			\multicolumn{1}{c|}{\multirow{2}{*}{\textbf{20}}} &
			\multicolumn{1}{l|}{\textit{\cellcolor[HTML]{EFEFEF}Execution time (sec)}} &
			\multicolumn{1}{l|}{\cellcolor[HTML]{EFEFEF}0.1} &
			\multicolumn{1}{l|}{\cellcolor[HTML]{EFEFEF}0.3s} &
			\multicolumn{1}{l|}{\cellcolor[HTML]{EFEFEF}0.5s} &
			\multicolumn{1}{l|}{\cellcolor[HTML]{EFEFEF}0.2s} &
			\multicolumn{1}{l|}{\cellcolor[HTML]{EFEFEF}1.1s} &
			\multicolumn{1}{l|}{\cellcolor[HTML]{EFEFEF}2s} &
			\multicolumn{1}{l|}{\cellcolor[HTML]{EFEFEF}0.5s} &
			\multicolumn{1}{l|}{\cellcolor[HTML]{EFEFEF}1.8s} &
			\multicolumn{1}{l|}{\cellcolor[HTML]{EFEFEF}3.8s} &
			\multicolumn{1}{l|}{\cellcolor[HTML]{EFEFEF}0.03s} \\ \cline{3-13} 
			\multicolumn{1}{|l|}{} &
			\multicolumn{1}{c|}{} &
			\multicolumn{1}{l|}{\textit{Overall loss}} &
			\multicolumn{1}{l|}{10/10} &
			\multicolumn{1}{l|}{10/10} &
			\multicolumn{1}{l|}{10/10} &
			\multicolumn{1}{l|}{10/10} &
			\multicolumn{1}{l|}{10/10} &
			\multicolumn{1}{l|}{10/10} &
			\multicolumn{1}{l|}{10/10} &
			\multicolumn{1}{l|}{10/10} &
			\multicolumn{1}{l|}{10/10} &
			\multicolumn{1}{l|}{\textbf{Success}} \\ \cline{2-13} 
			\multicolumn{1}{|l|}{} &
			\multicolumn{1}{c|}{\multirow{2}{*}{\textbf{50}}} &
			\multicolumn{1}{l|}{\textit{\cellcolor[HTML]{EFEFEF}Execution time (sec)}} &
			\multicolumn{1}{l|}{\cellcolor[HTML]{EFEFEF}0.1s} &
			\multicolumn{1}{l|}{\cellcolor[HTML]{EFEFEF}0.3s} &
			\multicolumn{1}{l|}{\cellcolor[HTML]{EFEFEF}0.6s} &
			\multicolumn{1}{l|}{\cellcolor[HTML]{EFEFEF}0.3s} &
			\multicolumn{1}{l|}{\cellcolor[HTML]{EFEFEF}1.2s} &
			\multicolumn{1}{l|}{\cellcolor[HTML]{EFEFEF}2.3s} &
			\multicolumn{1}{l|}{\cellcolor[HTML]{EFEFEF}0.5s} &
			\multicolumn{1}{l|}{\cellcolor[HTML]{EFEFEF}2.2s} &
			\multicolumn{1}{l|}{\cellcolor[HTML]{EFEFEF}4.2s} &
			\multicolumn{1}{l|}{\cellcolor[HTML]{EFEFEF}0.2s} \\ \cline{3-13} 
			\multicolumn{1}{|l|}{} &
			\multicolumn{1}{c|}{} &
			\multicolumn{1}{l|}{\textit{Overall loss}} &
			\multicolumn{1}{l|}{8/10} &
			\multicolumn{1}{l|}{10/10} &
			\multicolumn{1}{l|}{10/10} &
			\multicolumn{1}{l|}{10/10} &
			\multicolumn{1}{l|}{10/10} &
			\multicolumn{1}{l|}{10/10} &
			\multicolumn{1}{l|}{10/10} &
			\multicolumn{1}{l|}{10/10} &
			\multicolumn{1}{l|}{10/10} &
			\multicolumn{1}{l|}{\textbf{Failure}} \\ \cline{2-13} 
			\multicolumn{1}{|l|}{} &
			\multicolumn{1}{c|}{\multirow{2}{*}{\textbf{100}}} &
			\multicolumn{1}{l|}{\textit{\cellcolor[HTML]{EFEFEF}Execution time (sec)}} &
			\multicolumn{1}{l|}{\cellcolor[HTML]{EFEFEF}0.2s} &
			\multicolumn{1}{l|}{\cellcolor[HTML]{EFEFEF}0.7s} &
			\multicolumn{1}{l|}{\cellcolor[HTML]{EFEFEF}1.2s} &
			\multicolumn{1}{l|}{\cellcolor[HTML]{EFEFEF}0.7s} &
			\multicolumn{1}{l|}{\cellcolor[HTML]{EFEFEF}2.4s} &
			\multicolumn{1}{l|}{\cellcolor[HTML]{EFEFEF}4.6s} &
			\multicolumn{1}{l|}{\cellcolor[HTML]{EFEFEF}1.1s} &
			\multicolumn{1}{l|}{\cellcolor[HTML]{EFEFEF}4.5s} &
			\multicolumn{1}{l|}{\cellcolor[HTML]{EFEFEF}9s} &
			\multicolumn{1}{l|}{\cellcolor[HTML]{EFEFEF}1.3s} \\ \cline{3-13} 
			\multicolumn{1}{|l|}{} &
			\multicolumn{1}{c|}{} &
			\multicolumn{1}{l|}{\textit{Overall loss}} &
			\multicolumn{1}{l|}{7/10} &
			\multicolumn{1}{l|}{7/10} &
			\multicolumn{1}{l|}{10/10} &
			\multicolumn{1}{l|}{8/10} &
			\multicolumn{1}{l|}{10/10} &
			\multicolumn{1}{l|}{10/10} &
			\multicolumn{1}{l|}{9/10} &
			\multicolumn{1}{l|}{10/10} &
			\multicolumn{1}{l|}{10/10} &
			\multicolumn{1}{l|}{\textbf{Failure}} \\ \hline
			\multicolumn{1}{|l|}{\multirow{6}{*}{\textbf{\begin{tabular}[c]{@{}l@{}}Increasing number \\ of threats\\ (50 controls 40 GCD\\ with range of \\ 80 to 400 costs)\end{tabular}}}} &
			\multicolumn{1}{c|}{\multirow{2}{*}{\textbf{20}}} &
			\multicolumn{1}{l|}{\textit{\cellcolor[HTML]{EFEFEF}Execution time (sec)}} &
			\multicolumn{1}{l|}{\cellcolor[HTML]{EFEFEF}0.1s} &
			\multicolumn{1}{l|}{\cellcolor[HTML]{EFEFEF}0.3s} &
			\multicolumn{1}{l|}{\cellcolor[HTML]{EFEFEF}0.6s} &
			\multicolumn{1}{l|}{\cellcolor[HTML]{EFEFEF}0.3s} &
			\multicolumn{1}{l|}{\cellcolor[HTML]{EFEFEF}1.2s} &
			\multicolumn{1}{l|}{\cellcolor[HTML]{EFEFEF}2.3s} &
			\multicolumn{1}{l|}{\cellcolor[HTML]{EFEFEF}0.5s} &
			\multicolumn{1}{l|}{\cellcolor[HTML]{EFEFEF}2.2s} &
			\multicolumn{1}{l|}{\cellcolor[HTML]{EFEFEF}4.2s} &
			\multicolumn{1}{l|}{\cellcolor[HTML]{EFEFEF}0.2s} \\ \cline{3-13} 
			\multicolumn{1}{|l|}{} &
			\multicolumn{1}{c|}{} &
			\multicolumn{1}{l|}{\textit{Overall loss}} &
			\multicolumn{1}{l|}{8/10} &
			\multicolumn{1}{l|}{10/10} &
			\multicolumn{1}{l|}{10/10} &
			\multicolumn{1}{l|}{10/10} &
			\multicolumn{1}{l|}{10/10} &
			\multicolumn{1}{l|}{10/10} &
			\multicolumn{1}{l|}{10/10} &
			\multicolumn{1}{l|}{10/10} &
			\multicolumn{1}{l|}{10/10} &
			\multicolumn{1}{l|}{\textbf{Failure}} \\ \cline{2-13} 
			\multicolumn{1}{|l|}{} &
			\multicolumn{1}{c|}{\multirow{2}{*}{\textbf{30}}} &
			\multicolumn{1}{l|}{\textit{\cellcolor[HTML]{EFEFEF}Execution time (sec)}} &
			\multicolumn{1}{l|}{\cellcolor[HTML]{EFEFEF}0.2s} &
			\multicolumn{1}{l|}{\cellcolor[HTML]{EFEFEF}0.8s} &
			\multicolumn{1}{l|}{\cellcolor[HTML]{EFEFEF}1.4s} &
			\multicolumn{1}{l|}{\cellcolor[HTML]{EFEFEF}0.7s} &
			\multicolumn{1}{l|}{\cellcolor[HTML]{EFEFEF}2.6s} &
			\multicolumn{1}{l|}{\cellcolor[HTML]{EFEFEF}5.4s} &
			\multicolumn{1}{l|}{\cellcolor[HTML]{EFEFEF}1.2s} &
			\multicolumn{1}{l|}{\cellcolor[HTML]{EFEFEF}5.9s} &
			\multicolumn{1}{l|}{\cellcolor[HTML]{EFEFEF}10s} &
			\multicolumn{1}{l|}{\cellcolor[HTML]{EFEFEF}0.31s} \\ \cline{3-13} 
			\multicolumn{1}{|l|}{} &
			\multicolumn{1}{c|}{} &
			\multicolumn{1}{l|}{\textit{Overall loss}} &
			\multicolumn{1}{l|}{5/10} &
			\multicolumn{1}{l|}{10/10} &
			\multicolumn{1}{l|}{10/10} &
			\multicolumn{1}{l|}{8/10} &
			\multicolumn{1}{l|}{10/10} &
			\multicolumn{1}{l|}{10/10} &
			\multicolumn{1}{l|}{10/10} &
			\multicolumn{1}{l|}{10/10} &
			\multicolumn{1}{l|}{10/10} &
			\multicolumn{1}{l|}{\textbf{Failure}} \\ \cline{2-13} 
			\multicolumn{1}{|l|}{} &
			\multicolumn{1}{c|}{\multirow{2}{*}{\textbf{40}}} &
			\multicolumn{1}{l|}{\textit{\cellcolor[HTML]{EFEFEF}Execution time (sec)}} &
			\multicolumn{1}{l|}{\cellcolor[HTML]{EFEFEF}0.2s} &
			\multicolumn{1}{l|}{\cellcolor[HTML]{EFEFEF}1.1s} &
			\multicolumn{1}{l|}{\cellcolor[HTML]{EFEFEF}1.8s} &
			\multicolumn{1}{l|}{\cellcolor[HTML]{EFEFEF}0.9s} &
			\multicolumn{1}{l|}{\cellcolor[HTML]{EFEFEF}3.7s} &
			\multicolumn{1}{l|}{\cellcolor[HTML]{EFEFEF}7.2s} &
			\multicolumn{1}{l|}{\cellcolor[HTML]{EFEFEF}1.7s} &
			\multicolumn{1}{l|}{\cellcolor[HTML]{EFEFEF}7.3s} &
			\multicolumn{1}{l|}{\cellcolor[HTML]{EFEFEF}14s} &
			\multicolumn{1}{l|}{\cellcolor[HTML]{EFEFEF}0.38s} \\ \cline{3-13} 
			\multicolumn{1}{|l|}{} &
			\multicolumn{1}{c|}{} &
			\multicolumn{1}{l|}{\textit{Overall loss}} &
			\multicolumn{1}{l|}{2/10} &
			\multicolumn{1}{l|}{5/10} &
			\multicolumn{1}{l|}{8/10} &
			\multicolumn{1}{l|}{3/10} &
			\multicolumn{1}{l|}{9/10} &
			\multicolumn{1}{l|}{10/10} &
			\multicolumn{1}{l|}{8/10} &
			\multicolumn{1}{l|}{10/10} &
			\multicolumn{1}{l|}{10/10} &
			\multicolumn{1}{l|}{\textbf{Failure}} \\ \hline
			\multicolumn{1}{|l|}{\multirow{6}{*}{\textbf{\begin{tabular}[c]{@{}l@{}}Different GCD\\ (50 controls and \\ 20 threats with \\ range of 80 to \\ 400 costs)\end{tabular}}}} &
			\multicolumn{1}{c|}{\multirow{2}{*}{\textbf{40 GCD}}} &
			\multicolumn{1}{l|}{\textit{\cellcolor[HTML]{EFEFEF}Execution time (sec)}} &
			\multicolumn{1}{l|}{\cellcolor[HTML]{EFEFEF}0.1s} &
			\multicolumn{1}{l|}{\cellcolor[HTML]{EFEFEF}0.3s} &
			\multicolumn{1}{l|}{\cellcolor[HTML]{EFEFEF}0.6s} &
			\multicolumn{1}{l|}{\cellcolor[HTML]{EFEFEF}0.3s} &
			\multicolumn{1}{l|}{\cellcolor[HTML]{EFEFEF}1.2s} &
			\multicolumn{1}{l|}{\cellcolor[HTML]{EFEFEF}2.3s} &
			\multicolumn{1}{l|}{\cellcolor[HTML]{EFEFEF}0.5s} &
			\multicolumn{1}{l|}{\cellcolor[HTML]{EFEFEF}2.2s} &
			\multicolumn{1}{l|}{\cellcolor[HTML]{EFEFEF}4.2s} &
			\multicolumn{1}{l|}{\cellcolor[HTML]{EFEFEF}0.2s} \\ \cline{3-13} 
			\multicolumn{1}{|l|}{} &
			\multicolumn{1}{c|}{} &
			\multicolumn{1}{l|}{\textit{Overall loss}} &
			\multicolumn{1}{l|}{8/10} &
			\multicolumn{1}{l|}{10/10} &
			\multicolumn{1}{l|}{10/10} &
			\multicolumn{1}{l|}{10/10} &
			\multicolumn{1}{l|}{10/10} &
			\multicolumn{1}{l|}{10/10} &
			\multicolumn{1}{l|}{10/10} &
			\multicolumn{1}{l|}{10/10} &
			\multicolumn{1}{l|}{10/10} &
			\multicolumn{1}{l|}{\textbf{Failure}} \\ \cline{2-13} 
			\multicolumn{1}{|l|}{} &
			\multicolumn{1}{c|}{\multirow{2}{*}{\textbf{20 GCD}}} &
			\multicolumn{1}{l|}{\textit{\cellcolor[HTML]{EFEFEF}Execution time (sec)}} &
			\multicolumn{1}{l|}{\cellcolor[HTML]{EFEFEF}0.1s} &
			\multicolumn{1}{l|}{\cellcolor[HTML]{EFEFEF}0.3s} &
			\multicolumn{1}{l|}{\cellcolor[HTML]{EFEFEF}0.6s} &
			\multicolumn{1}{l|}{\cellcolor[HTML]{EFEFEF}0.3s} &
			\multicolumn{1}{l|}{\cellcolor[HTML]{EFEFEF}1.2s} &
			\multicolumn{1}{l|}{\cellcolor[HTML]{EFEFEF}2.3s} &
			\multicolumn{1}{l|}{\cellcolor[HTML]{EFEFEF}0.9s} &
			\multicolumn{1}{l|}{\cellcolor[HTML]{EFEFEF}3.8s} &
			\multicolumn{1}{l|}{\cellcolor[HTML]{EFEFEF}7.4s} &
			\multicolumn{1}{l|}{\cellcolor[HTML]{EFEFEF}0.2s} \\ \cline{3-13} 
			\multicolumn{1}{|l|}{} &
			\multicolumn{1}{c|}{} &
			\multicolumn{1}{l|}{\textit{Overall loss}} &
			\multicolumn{1}{l|}{5/10} &
			\multicolumn{1}{l|}{10/10} &
			\multicolumn{1}{l|}{10/10} &
			\multicolumn{1}{l|}{7/10} &
			\multicolumn{1}{l|}{10/10} &
			\multicolumn{1}{l|}{10/10} &
			\multicolumn{1}{l|}{9/10} &
			\multicolumn{1}{l|}{10/10} &
			\multicolumn{1}{l|}{10/10} &
			\multicolumn{1}{l|}{\textbf{Success}} \\ \cline{2-13} 
			\multicolumn{1}{|l|}{} &
			\multicolumn{1}{c|}{\multirow{2}{*}{\textbf{10 GCD}}} &
			\multicolumn{1}{l|}{\textit{\cellcolor[HTML]{EFEFEF}Execution time (sec)}} &
			\multicolumn{1}{l|}{\cellcolor[HTML]{EFEFEF}0.1s} &
			\multicolumn{1}{l|}{\cellcolor[HTML]{EFEFEF}0.3s} &
			\multicolumn{1}{l|}{\cellcolor[HTML]{EFEFEF}0.6s} &
			\multicolumn{1}{l|}{\cellcolor[HTML]{EFEFEF}0.3s} &
			\multicolumn{1}{l|}{\cellcolor[HTML]{EFEFEF}1.2s} &
			\multicolumn{1}{l|}{\cellcolor[HTML]{EFEFEF}2.1s} &
			\multicolumn{1}{l|}{\cellcolor[HTML]{EFEFEF}0.6s} &
			\multicolumn{1}{l|}{\cellcolor[HTML]{EFEFEF}2.2s} &
			\multicolumn{1}{l|}{\cellcolor[HTML]{EFEFEF}4.1s} &
			\multicolumn{1}{l|}{\cellcolor[HTML]{EFEFEF}0.21s} \\ \cline{3-13} 
			\multicolumn{1}{|l|}{} &
			\multicolumn{1}{c|}{} &
			\multicolumn{1}{l|}{\textit{Overall loss}} &
			\multicolumn{1}{l|}{5/10} &
			\multicolumn{1}{l|}{9/10} &
			\multicolumn{1}{l|}{10/10} &
			\multicolumn{1}{l|}{10/10} &
			\multicolumn{1}{l|}{10/10} &
			\multicolumn{1}{l|}{10/10} &
			\multicolumn{1}{l|}{10/10} &
			\multicolumn{1}{l|}{10/10} &
			\multicolumn{1}{l|}{10/10} &
			\multicolumn{1}{l|}{\textbf{Failure}} \\ \hline
			\multicolumn{1}{|l|}{\multirow{6}{*}{\textbf{\begin{tabular}[c]{@{}l@{}}Different range \\ of cost\\ (50 controls and \\ 20 threats with \\ 40 GCD)\end{tabular}}}} &
			\multicolumn{1}{c|}{\multirow{2}{*}{\textbf{\begin{tabular}[c]{@{}c@{}}range of \\ 80 - 400\end{tabular}}}} &
			\multicolumn{1}{l|}{\textit{\cellcolor[HTML]{EFEFEF}Execution time (sec)}} &
			\multicolumn{1}{l|}{\cellcolor[HTML]{EFEFEF}0.1s} &
			\multicolumn{1}{l|}{\cellcolor[HTML]{EFEFEF}0.3s} &
			\multicolumn{1}{l|}{\cellcolor[HTML]{EFEFEF}0.6s} &
			\multicolumn{1}{l|}{\cellcolor[HTML]{EFEFEF}0.3s} &
			\multicolumn{1}{l|}{\cellcolor[HTML]{EFEFEF}1.2s} &
			\multicolumn{1}{l|}{\cellcolor[HTML]{EFEFEF}2.3s} &
			\multicolumn{1}{l|}{\cellcolor[HTML]{EFEFEF}0.5s} &
			\multicolumn{1}{l|}{\cellcolor[HTML]{EFEFEF}2.2s} &
			\multicolumn{1}{l|}{\cellcolor[HTML]{EFEFEF}4.2s} &
			\multicolumn{1}{l|}{\cellcolor[HTML]{EFEFEF}0.2s} \\ \cline{3-13} 
			\multicolumn{1}{|l|}{} &
			\multicolumn{1}{c|}{} &
			\multicolumn{1}{l|}{\textit{Overall loss}} &
			\multicolumn{1}{l|}{8/10} &
			\multicolumn{1}{l|}{10/10} &
			\multicolumn{1}{l|}{10/10} &
			\multicolumn{1}{l|}{10/10} &
			\multicolumn{1}{l|}{10/10} &
			\multicolumn{1}{l|}{10/10} &
			\multicolumn{1}{l|}{10/10} &
			\multicolumn{1}{l|}{10/10} &
			\multicolumn{1}{l|}{10/10} &
			\multicolumn{1}{l|}{\textbf{Failure}} \\ \cline{2-13} 
			\multicolumn{1}{|l|}{} &
			\multicolumn{1}{c|}{\multirow{2}{*}{\textbf{\begin{tabular}[c]{@{}c@{}}range of \\ 80 - 240\end{tabular}}}} &
			\multicolumn{1}{l|}{\textit{\cellcolor[HTML]{EFEFEF}Execution time (sec)}} &
			\multicolumn{1}{l|}{\cellcolor[HTML]{EFEFEF}0.1s} &
			\multicolumn{1}{l|}{\cellcolor[HTML]{EFEFEF}0.3s} &
			\multicolumn{1}{l|}{\cellcolor[HTML]{EFEFEF}0.6s} &
			\multicolumn{1}{l|}{\cellcolor[HTML]{EFEFEF}0.3s} &
			\multicolumn{1}{l|}{\cellcolor[HTML]{EFEFEF}1.2s} &
			\multicolumn{1}{l|}{\cellcolor[HTML]{EFEFEF}2.2s} &
			\multicolumn{1}{l|}{\cellcolor[HTML]{EFEFEF}0.5s} &
			\multicolumn{1}{l|}{\cellcolor[HTML]{EFEFEF}2.2s} &
			\multicolumn{1}{l|}{\cellcolor[HTML]{EFEFEF}4.3s} &
			\multicolumn{1}{l|}{\cellcolor[HTML]{EFEFEF}0.2s} \\ \cline{3-13} 
			\multicolumn{1}{|l|}{} &
			\multicolumn{1}{c|}{} &
			\multicolumn{1}{l|}{\textit{Overall loss}} &
			\multicolumn{1}{l|}{3/10} &
			\multicolumn{1}{l|}{8/10} &
			\multicolumn{1}{l|}{8/10} &
			\multicolumn{1}{l|}{5/10} &
			\multicolumn{1}{l|}{7/10} &
			\multicolumn{1}{l|}{8/10} &
			\multicolumn{1}{l|}{6/10} &
			\multicolumn{1}{l|}{10/10} &
			\multicolumn{1}{l|}{10/10} &
			\multicolumn{1}{l|}{\textbf{Failure}} \\ \cline{2-13} 
			\multicolumn{1}{|l|}{} &
			\multicolumn{1}{c|}{\multirow{2}{*}{\textbf{\begin{tabular}[c]{@{}c@{}}range of \\ 80 - 160\end{tabular}}}} &
			\multicolumn{1}{l|}{\textit{\cellcolor[HTML]{EFEFEF}Execution time (sec)}} &
			\multicolumn{1}{l|}{\cellcolor[HTML]{EFEFEF}0.1s} &
			\multicolumn{1}{l|}{\cellcolor[HTML]{EFEFEF}0.3s} &
			\multicolumn{1}{l|}{\cellcolor[HTML]{EFEFEF}0.6s} &
			\multicolumn{1}{l|}{\cellcolor[HTML]{EFEFEF}0.3s} &
			\multicolumn{1}{l|}{\cellcolor[HTML]{EFEFEF}1.2s} &
			\multicolumn{1}{l|}{\cellcolor[HTML]{EFEFEF}2.2s} &
			\multicolumn{1}{l|}{\cellcolor[HTML]{EFEFEF}0.5s} &
			\multicolumn{1}{l|}{\cellcolor[HTML]{EFEFEF}2.3s} &
			\multicolumn{1}{l|}{\cellcolor[HTML]{EFEFEF}4.2s} &
			\multicolumn{1}{l|}{\cellcolor[HTML]{EFEFEF}0.2s} \\ \cline{3-13} 
			\multicolumn{1}{|l|}{} &
			\multicolumn{1}{c|}{} &
			\multicolumn{1}{l|}{\textit{Overall loss}} &
			\multicolumn{1}{l|}{3/10} &
			\multicolumn{1}{l|}{4/10} &
			\multicolumn{1}{l|}{7/10} &
			\multicolumn{1}{l|}{5/10} &
			\multicolumn{1}{l|}{5/10} &
			\multicolumn{1}{l|}{8/10} &
			\multicolumn{1}{l|}{8/10} &
			\multicolumn{1}{l|}{10/10} &
			\multicolumn{1}{l|}{10/10} &
			\multicolumn{1}{l|}{\textbf{Failure}} \\ \hline
			\multicolumn{1}{|l|}{\multirow{6}{*}{\textbf{\begin{tabular}[c]{@{}l@{}}Affected threats\\ (50 controls and \\ 20 threats with \\ range of 80 to 400 \\ costs and 40 GCD)\end{tabular}}}} &
			\multicolumn{1}{c|}{\multirow{2}{*}{\textbf{\begin{tabular}[c]{@{}c@{}}1 threat \\ only\end{tabular}}}} &
			\multicolumn{1}{l|}{\textit{\cellcolor[HTML]{EFEFEF}Execution time (sec)}} &
			\multicolumn{1}{l|}{\cellcolor[HTML]{EFEFEF}0.2s} &
			\multicolumn{1}{l|}{\cellcolor[HTML]{EFEFEF}0.8s} &
			\multicolumn{1}{l|}{\cellcolor[HTML]{EFEFEF}1.5s} &
			\multicolumn{1}{l|}{\cellcolor[HTML]{EFEFEF}0.7s} &
			\multicolumn{1}{l|}{\cellcolor[HTML]{EFEFEF}3.1s} &
			\multicolumn{1}{l|}{\cellcolor[HTML]{EFEFEF}6s} &
			\multicolumn{1}{l|}{\cellcolor[HTML]{EFEFEF}1.3s} &
			\multicolumn{1}{l|}{\cellcolor[HTML]{EFEFEF}6.1s} &
			\multicolumn{1}{l|}{\cellcolor[HTML]{EFEFEF}22s} &
			\multicolumn{1}{l|}{\cellcolor[HTML]{EFEFEF}0.03s} \\ \cline{3-13} 
			\multicolumn{1}{|l|}{} &
			\multicolumn{1}{c|}{} &
			\multicolumn{1}{l|}{\textit{Overall loss}} &
			\multicolumn{1}{l|}{10/10} &
			\multicolumn{1}{l|}{10/10} &
			\multicolumn{1}{l|}{10/10} &
			\multicolumn{1}{l|}{10/10} &
			\multicolumn{1}{l|}{10/10} &
			\multicolumn{1}{l|}{10/10} &
			\multicolumn{1}{l|}{10/10} &
			\multicolumn{1}{l|}{10/10} &
			\multicolumn{1}{l|}{10/10} &
			\multicolumn{1}{l|}{\textbf{Success}} \\ \cline{2-13} 
			\multicolumn{1}{|l|}{} &
			\multicolumn{1}{c|}{\multirow{2}{*}{\textbf{\begin{tabular}[c]{@{}c@{}}3 threats \\ affected\end{tabular}}}} &
			\multicolumn{1}{l|}{\textit{\cellcolor[HTML]{EFEFEF}Execution time (sec)}} &
			\multicolumn{1}{l|}{\cellcolor[HTML]{EFEFEF}0.2s} &
			\multicolumn{1}{l|}{\cellcolor[HTML]{EFEFEF}0.6s} &
			\multicolumn{1}{l|}{\cellcolor[HTML]{EFEFEF}1.3s} &
			\multicolumn{1}{l|}{\cellcolor[HTML]{EFEFEF}0.6s} &
			\multicolumn{1}{l|}{\cellcolor[HTML]{EFEFEF}2.8s} &
			\multicolumn{1}{l|}{\cellcolor[HTML]{EFEFEF}5s} &
			\multicolumn{1}{l|}{\cellcolor[HTML]{EFEFEF}1.1s} &
			\multicolumn{1}{l|}{\cellcolor[HTML]{EFEFEF}5.1s} &
			\multicolumn{1}{l|}{\cellcolor[HTML]{EFEFEF}18s} &
			\multicolumn{1}{l|}{\cellcolor[HTML]{EFEFEF}0.06s} \\ \cline{3-13} 
			\multicolumn{1}{|l|}{} &
			\multicolumn{1}{c|}{} &
			\multicolumn{1}{l|}{\textit{Overall loss}} &
			\multicolumn{1}{l|}{5/10} &
			\multicolumn{1}{l|}{10/10} &
			\multicolumn{1}{l|}{10/10} &
			\multicolumn{1}{l|}{10/10} &
			\multicolumn{1}{l|}{10/10} &
			\multicolumn{1}{l|}{10/10} &
			\multicolumn{1}{l|}{10/10} &
			\multicolumn{1}{l|}{10/10} &
			\multicolumn{1}{l|}{10/10} &
			\multicolumn{1}{l|}{\textbf{Failure}} \\ \cline{2-13} 
			\multicolumn{1}{|l|}{} &
			\multicolumn{1}{c|}{\multirow{2}{*}{\textbf{\begin{tabular}[c]{@{}c@{}}7 threats \\ affected\end{tabular}}}} &
			\multicolumn{1}{l|}{\textit{\cellcolor[HTML]{EFEFEF}Execution time (sec)}} &
			\multicolumn{1}{l|}{\cellcolor[HTML]{EFEFEF}0.1s} &
			\multicolumn{1}{l|}{\cellcolor[HTML]{EFEFEF}0.5s} &
			\multicolumn{1}{l|}{\cellcolor[HTML]{EFEFEF}0.9s} &
			\multicolumn{1}{l|}{\cellcolor[HTML]{EFEFEF}0.5s} &
			\multicolumn{1}{l|}{\cellcolor[HTML]{EFEFEF}1.9s} &
			\multicolumn{1}{l|}{\cellcolor[HTML]{EFEFEF}3.6s} &
			\multicolumn{1}{l|}{\cellcolor[HTML]{EFEFEF}0.9s} &
			\multicolumn{1}{l|}{\cellcolor[HTML]{EFEFEF}3.5s} &
			\multicolumn{1}{l|}{\cellcolor[HTML]{EFEFEF}7s} &
			\multicolumn{1}{l|}{\cellcolor[HTML]{EFEFEF}0.09s} \\ \cline{3-13} 
			\multicolumn{1}{|l|}{} &
			\multicolumn{1}{c|}{} &
			\multicolumn{1}{l|}{\textit{Overall loss}} &
			\multicolumn{1}{l|}{4/10} &
			\multicolumn{1}{l|}{7/10} &
			\multicolumn{1}{l|}{10/10} &
			\multicolumn{1}{l|}{6/10} &
			\multicolumn{1}{l|}{10/10} &
			\multicolumn{1}{l|}{10/10} &
			\multicolumn{1}{l|}{8/10} &
			\multicolumn{1}{l|}{10/10} &
			\multicolumn{1}{l|}{10/10} &
			\multicolumn{1}{l|}{\textbf{Failure}} \\ \hline
		\end{tabular}%
	}
	\caption{Different configurations for GA solution and their comparison with Greedy}
	\label{tab:my-table}
\end{table}
Table~\ref{tab:my-table} shows the result of different limits and population sizes for GA solution and their comparison with the Greedy results (binary representation of either \emph{Success} or \emph{Failure} for finding the optimal answer). Every experiment for GA settings has been conducted 10 times to evaluate the accuracy of the solution and the average execution time (highlighted and represented in seconds) is reported. 

It is conceivable that with small population size and limit, GA performs much faster but fails more often. Also, other factors, i.e., GCD and Range of costs, affect the result of GA. Depending on the limit and population size as well as other factors, the gap between results is not huge. For instance, when we have a range of 80 - 160 costs with 100 limit and 100 population size, we see 4 different outcomes which difference are 29.4 (about 5\% difference with the optimal answer) utmost. 

The Greedy solution yields the result in a relatively shorter time as expected, yet fails to find the optimal solution in almost every experiment. 

\subsection{Analysis of results}
Our experiments show that our DP solution with projection really improves the speed of searching for the answer. The algorithm produces the results in a matter of 5-10 minutes for about 10-20 threats and for 50-100 controls. Since the decisions to be taken are strategic, i.e., will be valid for a long time (e.g., a year) we may see this time as acceptable. Moreover, the usage of a more powerful computing system will increase the speed of the proposed solutions. Furthermore, advanced methods for computation (e.g., parallel computations) will improve the speed even more.  

Obviously, the exact algorithm cannot beat the approximate ones (although, we see that for a smaller quantity of input parameters, DP-based solutions are even faster than GA algorithms), but this is the price for assurance in the exact answer the algorithm produces. Naturally, if one considers the full set of controls from a standard (e.g., NIST \cite{NIST800} or ISO-27002 \cite{ISO}) and the full set of threats (e.g., see ISO 27005 \cite{ISO27005}) it is better to use GA. On the other hand, an organisation thinking about improving its self-protection is most likely to consider a moderate number of possible controls (i.e., 10-20) and be more focused on a moderate number of relevant threats (i.e., 10-20), thus, making application of our DP-based solution acceptable. Naturally, if the numbers are very small, an exhaustive search may be also applicable, but its time of execution grows too fast comparing to dynamic programming based algorithms.

Last, but not least, not only quantity but also the quality of the input has a great impact on the algorithms. Once GCD of the costs for security controls is high enough and their range is large, the DP-based solutions are able to terminate in a reasonable time. However, the reverse cases lead to much longer time to terminate. We also see that once only a few controls are affected, our solution with projection idea significantly outperforms the legacy one.  

In sum, we may conclude that once a company has to select controls out of a limited number of alternatives, our solution is applicable in reality. On the other hand, with a larger number of controls, GA is more appropriate. 

%% file: Discussion.tex
\section{Discussion}
\label{sec:MAinDiscussion}
In this section we would like to discuss some limitations of our work, most of which relate to its practical application. Our solution is based on the competitive insurance market model, which is simple and widely used in theoretical insurance studies \cite{Rothschild, Akerlof}. This model allows us to focus on the core problems and simplifies the analysis. On the other hand, we acknowledge that in reality such a model should be adjusted to take into account various features deviating the result from the theoretical analysis. In case of insurance, such deviation is often modelled with additional loading factors \cite{Marotta2017}, which increases the price to cover insurance expenses, ensures solvency, securing insurance risks, etc. In this paper, we do not consider how much the non-competitive market aspects affect the results of our analysis and leave this issue for future studies. We also would like to underline that other factors (e.g., insurance regulations, deducible, perception of risk, the interdependence of cyber security, etc.)  may affect the results of our study and require specific analysis to estimate the deviation from our core model.  

In this work, we assumed security controls' costs and probabilities to be independent of each other. In practice, some controls may require installation of others (e.g., security audit may require the presence of monitoring and logging mechanisms) or even conflict with each other (e.g., cryptography may reduce the effectiveness of security audit). Some of these issues can be resolved by pre-processing the input data (e.g., grouping some controls) and others can be resolved by some adjustments in the core algorithms. We leave these steps for future work.

Last but not least, we admit that our proposal relies on the knowledge of survival probabilities and expected losses, the values which are hard to find. The problem of identifying these values is well-known in the cyber security research and industrial community and is heavily related to the lack of statistical data available for research which is only becoming available \cite{Roman18}. On the other hand, we believe such values can be found by cyber insurers who are collecting data from their customers and are interested to use them for identification of these values. We also may see some real steps in this direction as IBM Security and Ponemon institute's Data Breach report \cite{Ponemon} contains the study (and reports statistics) for the loss estimation depending on the security controls installed.

%% file: 07_Conclusion.tex
\section{Conclusion and Future work}
\label{sec:Discussion}

In this work, we theoretically analysed the problem of investments distributions for a risk-averse model of an organisation which is considering to buy cyber insurance. We have found that even in presence of several threats at the same time a competitive insurance market leads to full insurance coverage as optimal.

Also, our approach helps the organisation to identify the security controls required to be installed to optimise investments in self-protection and cyber insurance. The final problem we were solving aims for a cost-effective set of controls, in contrast to the usual state of the art approach aiming at selecting the controls with the total cost fitting the budget (and, thus, which could be not cost-efficient). 

We provided an algorithmic solution which returns the exact answer to the main problem. The proposed solution proves to be faster than the direct application of the legacy approach our solution is based on. The solution is suitable for analysis of a reasonable set of available alternative controls (about 10-40) and threats (about 10), but for much larger sets of controls and threats genetic algorithm are more advisable to use. Although GA sometimes returns near optimal solutions, it is much more reliable than the Greedy algorithm and reasonably fast even with high settings. 

\section{Acknowledgement}
\label{sec:Ack}
This work was partially supported by projects H2020 MSCA NeCS 675320, H2020 MSCA CyberSure 734815, H2020 SPARTA 830892 and H2020 CyberSec4Europe 830929. We thank to anonymous reviewers.

%% file: 99_Apendix.tex
\section{Appendix}
\label{sec:appendix}

\subsection{Indemnity}
\label{sec:Indemnity}
Although it has already been proven in the literature that for a competitive market the optimal indemnity is equal to loss \cite{Marotta2017}, we need to prove this also for a multi-threat scenario. In this regard, we apply Jensen's inequality for a concave function (for any concave function $\phi(t)$ $E[\phi(t)]\leq \phi(E[t])$) for Equation~\ref{eq:ExpectedUtility}:
\begin{align}
&\sum_{\forall \vec{z}}p(\vec{z}|K_{s},x) \cdot U(W^0-(\vec{F}\odot \vec{p}(K_{s}|x))\times \vec{I}-x+\vec{z}\times(\vec{I}-\vec{L})) \leq \nonumber \\
&U(\sum_{\forall \vec{z}}p(\vec{z}|K_{s},x)\cdot \left[ W^0-(\vec{F}\odot \vec{p}(K_{s}|x))\times \vec{I}-x+\vec{z}\times(\vec{I}-\vec{L})\right]) = \nonumber \\
&U(\left[\sum_{\forall \vec{z}} p(\vec{z}|K_{s},x)\right](W^0-x)-\left[\sum_{\forall \vec{z}} p(\vec{z}|K_{s},x)\right] \cdot \left[ (\vec{F}\odot \vec{p}(K_{s}|x))\times \vec{I}\right]+ \nonumber \\
&\left[\sum_{\forall \vec{z}} p(\vec{z}|K_{s},x)\cdot \vec{z}\right]\times\vec{I}-\left[\sum_{\forall \vec{z}} p(\vec{z}|K_{s},x)\cdot\vec{z}\right]\times\vec{L}). \nonumber
\end{align}
Since $\sum_{\forall \vec{z}} p(\vec{z}|K_{s},x)=1$ and $\vec{F}\odot \vec{p}(K_{s}|x)=\sum_{\forall \vec{z}}p(\vec{z}|K_{s},x)\cdot \vec{z}$, we get:
\begin{align}
&U(W^0-x -\left[ (\vec{F}\odot \vec{p}(K_{s}|x))\times \vec{I}\right]+ (\vec{F}\odot \vec{p}(K_{s}|x))\times\vec{I}-(\vec{F}\odot \vec{p}(K_{s}|x))\times\vec{L})= \nonumber\\
&U(W^0-x -(\vec{F}\odot \vec{p}(K_{s}|x))\times\vec{L}). \nonumber
\end{align}
The last part ($U(W^0-x -(\vec{F}\odot \vec{p}(K_{s}|x))\times\vec{L})$) is the expected utility if $\vec{I}=\vec{L}$. In other words, Equation~\ref{eq:ExpectedUtility} is maximal if $\vec{I}=\vec{L}$. 

\subsection{Algorithms}
\label{app:algos}
\subsubsection{Backtracking Algorithm}
\begin{algorithm}[H]
\DontPrintSemicolon
\KwFunction{BackTrack}\;
\KwInput{$x^*,~ c, ~T,~C$}
\KwRequire{}
$x^*$ \tcp*{optimal investment found by \textit{searchForOptimalInvestments} algorithm} 
$c$ \tcp*{cost of controls}
$T$ \tcp*{auxiliary matrix T from \textit{searchForOptimalInvestments} algorithm}
$C$ \tcp*{CGD for control cost}
\KwEnsure{Optimal $K_s$}
$ K_s := \emptyset$ \;
$x:=x^*$\tcp*{we start with optimal investment level}
\For{i := 0 to n}
{ \tcp{Iteration starts from the end of control's list}
\If{$x-c[n-i-1] < 0$} 
{ 
\If{$T[n - i - 1][x/C] ~\neq~ T[n - i][x/C]$}
{
$ K_s $.append(k[n - i]) \tcp*{add $n-i$ -th control the selected controls list}
$x$ := $x$ - c[n - i - 1] \tcp*{Decrease the optimal investment by the cost of (n-i-1)-th control}
}
}	
}
\KwReturn{$ K_s$} 
\caption{Recover selected controls }\label{alg:ext}
\end{algorithm}
\medskip

\subsubsection{Greedy Algorithm}
\label{sec:GreAlg}
Function \emph{GreedySelection} (lines \ref{Gre6} to \ref{Gre7}) in algorithm \ref{alg:OptSelFunc} is the core function of the algorithm. It starts with an assumption that all controls are to be selected and gradually remove the ones which removal reduces the overall expenditure more than others. It stops when there is no control which removal decreases the expenditure. Auxiliary Function \emph{Calc} (Algorithm~\ref{alg:CalcFunc}) simply computes the overall expenditure with a set of currently selected controls; and auxiliary function \emph{FindWorstControl} (Algorithm~\ref{alg:UselessFunc}) selects the worst control to be removed.

\begin{algorithm}[H]
	\DontPrintSemicolon
	
	\KwFunction{GreedySelection} \label{Gre6} \;
	\KwInput{$c$, $\pi$, $\vec{L}$, $\vec{p}_{init}$, $\vec{F}$}
	\KwEnsure{lowest \textit{$minCost$} and \textit{$optimalBudget$} for optimal security investment $ x^* $}
	minCost, optimalBudget := \texttt{\textit{Calc($c$, $\pi$, $\vec{L}$, $\vec{p}_{init}$, $\vec{F}$)}} \tcp*{call \textit{Calc} function to \textit{minCost} and \textit{optimalBudget}}
	index := \texttt{\textit{FindWorstControl($c$, $\pi$, $\vec{L}$, $\vec{p}_{init}$, $\vec{F}$)}}\;
	\While{ index $\neq$ -1}
	{ \tcp{iterate until the loop ends}
	delete c[index] \tcp*{Delete the cost of the worst control}
	delete $\pi$[index] \tcp*{Delete all probabilities of survival related to the worst control}
	currentCost, currentBudget := \texttt{\textit{Calc($c$, $\pi$, $\vec{L}$, $\vec{p}_{init}$, $\vec{F}$)} \tcp*{call \textit{Calc} function to \textit{currentCost} and \textit{currentBudget}}
	\If{currentCost $ < $ minCost} { \tcp*{check the condition and replace the \textit{minCost} with \textit{currentCost} if it meets}\label{Gre8}
		{minCost} := {currentCost}\;
		{optimalBudget} := {currentBudget}\;
		}
	}
	index := \texttt{\textit{FindWorstControl($c$, $\pi$, $\vec{L}$, $\vec{p}_{init}$, $\vec{F}$)}}\;
	}
\Return{minCost, optimalBudget} \label{Gre7}

	\caption{The main function for Greedy approach}	\label{alg:OptSelFunc}	
\end{algorithm}
\medskip
\begin{algorithm}[H]
	\DontPrintSemicolon	
	\KwFunction{Calc} \label{Gre1} \tcp*{This function computes the \textit{minCost}(expenditure)}
	\KwInput{$c$, $\pi$, $\vec{L}$, $\vec{p}_{init}$, $\vec{F}$}
	\textit{minCost} := 0\;
	\For{ i := 1 \textbf{to} length($\vec{L}$)}
	{
	prob := 1 \tcp*{expected loss per threat i}
	\For{j := 1 \textbf{to} length(c)}
	{
	prob := $prob \cdot \pi[j][i]$
	}
	\textit{minCost} := $\vec{F}[i]\cdot\vec{L}[i] \cdot \vec{p}_{init}[i]\cdot prob$ \tcp*{put the sum of expected losses to the \textit{minCost}}
	}
	\textit{optimalBudget} := sum(c) \tcp*{put sum of costs to \textit{opimalBudget}}
	\textit{minCost} := \textit{minCost} + \textit{optimalBudget} \tcp*{computes the overall expenditure}
	
	\Return{minCost, optimalBudget} \label{Gre1}
	
	\caption{Calculation function of Greedy approach}	\label{alg:CalcFunc}
\end{algorithm}
\medskip
\begin{algorithm}[H]
	\DontPrintSemicolon
	\KwFunction{FindWorstControl} \label{Gre3} \tcp*{function for finding   worst controls}
	\KwInput{$c$, $\pi$, $\vec{L}$, $\vec{p}_{init}$, $\vec{F}$}
	\KwEnsure{Index of a control with the smallest risk reduction}
	\textit{n} := length(c) \tcp*{define the length of costs}
	\If {$ n == 1 $}
	{
		\Return{$-1$}
	}
	\textit{h} := length($\vec{L}$)\tcp*{define the length of probability of survival}
	\textit{s} := sum(c) \tcp*{define the length of overall cost}
	\textit{minCost}, \textit{temp} := \texttt{\textit{Calc}}($c$, $\pi$, $\vec{L}$, $\vec{p}_{init}$, $\vec{F}$) \tcp*{call the \textit{Calc()} function}
	\textit{index} := -1\;
	\For{j := 1 \textbf{to} n}    
	{
	$ val := 0 $\tcp*{the residual risk without j-th control}
	\For{i := 1 \textbf{to} h}{
        prob := 1\tcp*{probability of survival of threat i}
		\For{l := 1 to n}{
		\If{$l \neq j$}{
		$prob$ := $prob \cdot \pi[j][i]$ \tcp*{add the relation of $\pi$}
		}
		$val := val+\vec{F}[i] \cdot \vec{L}[i] \cdot \vec{p}_{init}[i]\cdot prob$
		}
		}
		
		\textit{currentCost} := $val$ + s - c[i] \tcp*{compute the \textit{currentCost}}
		\If {$ currentCost < minCost $}	
		{ \label{Gre5}
			\textit{minCost} := \textit{currentCost} \tcp*{update the \textit{minCost}}
			\textit{index} := i \tcp*{take the i-th control's index and return}
		}
	}
	\Return{index}\; \label{Gre4}
	\caption{Removal of the worst control for Greedy approach}	\label{alg:UselessFunc}
\end{algorithm}
%\medskip
\subsubsection{Genetic Algorithm}
\label{sec:GAlg}
\begin{algorithm}
	\DontPrintSemicolon
	\KwFunction{GASelection}\;
	\KwInput{$K$, $ \pi $, $\vec{L}$, $\vec{F}$, $ LIMIT $}
	\KwRequire{}
	$LIMIT \in \mathbb{N}$ \tcp*{a constant value to run the GA}
	\KwEnsure{\textit{lowest} $ expenditure := (\vec{F}\odot\prod \pi(K_{s}|x))\times \vec{L} + x$}
	generateRandomChromosomes(\textit{chromosomes}, $POP_{Num}$) \tcp*{generate initial population of chromosomes}
	sortByExpenditure(chromosomes)\tcp*{sort chromosomes by expenditure from lowest to largest}
	\textit{theAnswer} := \textit{chromosomes}[0] \tcp*{Initializing \textit{theAnswer}}
	\For{$ i := 1 ~ \textbf{to}~  LIMIT $}
	{
	    \textit{chromosomes} := \textit{CrossOver(\textit{nextGenChromosomes})}
        \tcp*{Call CrossOver}
        \If{\textit{theAnswer.expenditure} $<$ \textit{chromosomes}[0].expenditure}
        { 
        theAnswer := chromosomes[0]\tcp*{ensuring the best chromosome}
        }
   	    \textit{chromosomes} := \textit{Mutation(\textit{nextGenChromosomes})} \tcp*{Call Mutation Function}
   	            \If{\textit{theAnswer.expenditure} $<$ \textit{chromosomes}[0].expenditure}
        {
        theAnswer := chromosomes[0]\tcp*{update the best chromosome}
        }
	}
	\KwReturn{\textit{theAnswer}}

	\caption{The main function of the GA algorithm}\label{alg:MainFunc}
\end{algorithm}
\medskip
\begin{algorithm}
	\DontPrintSemicolon
	\KwFunction{Merge}\;
	\KwInput{ch1, ch2, Y1, $n_k$}
	\KwRequire{}
	$Y1 \in \mathbb{N}$ \tcp*{percentage of two-point CrossOver technique}
	$\textit{ch1}$ \tcp*{a chromosome in a pair}
	$\textit{ch2}$ \tcp*{another chromosome in a pair}
	$n_k \in \mathbb{N}$ \tcp*{number of controls/genes in the chromosome}
	\KwEnsure{Merging two chromosomes ch1 and ch2}
	y = rand.range(0, 100) \label{new2} \tcp*{some random percentage} 
	\If{$ y <= Y1 $} 
	{ \label{new3}
	l := rand.range(0, length(ch1.genes)-1)\;
	r := rand.range(l, length(ch1.genes)-1)\;

		\For{i := l \textbf{to} r}
		{
		ch1.genes[i] $\leftrightarrow$ ch2.genes[i]\tcp*{swap i-th genes of two chromosomes}
		}
	}
	\Else
	{
    	\For{i:=0 \textbf{to} $n_k$/2}
    	{
    	\label{new4}ch1.genes[i] $\leftrightarrow$ ch2.genes[i]
    	\tcp*{swap i-th genes of two chromosomes} 
    	}
		
	}
	\Return{ch1, ch2}
	\tcp*{Return a pair of chromosomes}
	\caption{Merge Function for the GA}\label{alg:Merge}
\end{algorithm}
\medskip
\begin{algorithm}
	\DontPrintSemicolon
	\KwFunction{Crossover}\;
	\KwInput{\textit{chromosomes}, $x$, $ N $, $ LIMIT $, $Y2$, $Y3$, $b1$, $b2$}
	\KwRequire{}
	\textit{chromosomes} \tcp*{a list of current chromosomes}
	$N \in \mathbb{N}$ \tcp*{number of chromosomes in the population}
	$b1 \in \mathbb{N}$ \tcp*{crossover percentage of good and good chromosomes}
	$b2 \in \mathbb{N}$ \tcp*{crossover percentage of good and bad chromosomes}
	$Y2 \in \mathbb{N}$ \tcp*{percentage of chromosomes for Elitism techniques}
	$Y3 \in \mathbb{N}$ \tcp*{percentage of good chromosomes of the population}
	\KwEnsure{\textit{next} \textit{Generation} of \textit{Chromosomes} obtained by CrossOver}
	\textit{nextGenChromosomes} := [] \tcp*{Define \textit{nextGenChromosomes} empty list}
	
	E := $ Y2 \cdot N / 100$ \label{new7}\tcp*{Define the percentage of elitism in integer}
	
	\For{$ i := 1 ~ \textbf{to}~  E $}
	{
		\textit{nextGenChromosomes}.append(chromosomes[i]) \tcp*{Add i-th chromosome without doing crossover} \label{new8}
	}
    r := $Y3 \cdot N / 100$ \tcp*{turn the percentage of good chromosomes into integer}
		\For{i := E+1 \textbf{to} N}
		{
		y := rand.range(0,100) \tcp*{define a random integer}
		\If{$ y <= b1 $}
		{ \tcp*{check if the crossover percentage of good chromosomes is greater than than a randomly chosen y}
			ii := rand.range(0, r) \tcp*{Define a random range of good chromosomes)}
			jj := rand.range(0, r)\;
		}
		\ElseIf{$ y <= b1 + b2 $}
		{ \tcp*{check if the crossover percentage of good and bad chromosomes is greater than than a randomly chosen y}
			ii := rand.range(0, r)\;
			jj := rand.range(r+1, N-1) \tcp*{Random range of bad chromosomes}
		}
		\Else
		{ \tcp*{check if the crossover percentage of bad and bad chromosomes is greater than than a randomly chosen y}
			ii := rand.range(r+1, N-1)\;
			jj := rand.range(r+1, N-1)\;
		}
		ch1, ch2 := Merge(chromosomes[ii], chromosomes[jj])\;
		\If{$ ch1.cost < x~\textbf{or}~x = 0 $}
		{
		\tcp*{check if cost of ch1 is less than investment or equal to $0$}
			\textit{nextGenChromsomes}.append(ch1)\;
		}
		\If{$ ch2.cost < x~\textbf{or}~x = 0 $}
		{ 
		\tcp*{check if cost of ch1 is less than investment or equal to $0$}
		    \textit{nextGenChromsomes}.append(ch2)\; \label{new9}
		}
		}
	sort(\textit{nextGenChromosomes}) \tcp*{\textbf{Sort} the chromosomes by \textit{expenditure}}
	\KwReturn{\textit{nextGenChromosomes}}
	\caption{CrossOver function for the New Population of the GA}\label{alg:NewPop}
\end{algorithm}
\medskip
\begin{algorithm}
	\DontPrintSemicolon
	\KwFunction{Mutation}\;
	\KwInput{\textit{chromosomes}, $ Y2 $, $b3$}
	\textit{chromosomes} \tcp*{a list of current chromosomes}
	$N \in \mathbb{N}$ \tcp*{number of chromosomes in the population}
	$Y2 \in \mathbb{N}$ \tcp*{percentage of chromosomes for Elitism techniques}
    $b3 \in \mathbb{N}$ \tcp*{defined number of bits to mutate}
    \KwEnsure{\textit{next} \textit{Generation} of \textit{Chromosomes} (\textit{nextGenChromosomes}) obtained by Mutation}
    \textit{nextGenChromosomes} := [] \tcp*{Define \textit{nextGenChromosomes} empty list}
	E := $Y2 \cdot N / 100$ \label{new7}\tcp*{Define the percentage of elitism as integer}
	\For{$ i := 1 ~ $\textbf{to}$~  E $}
	{
		\textit{nextGenChromosomes}.append(\textit{chromosomes}[i]) \tcp*{update the list} \label{new8}
	}
	\For{$ i := E + 1~ \textbf{to}~ N $}
	{
	    ch := \textit{chromosomes}[i]\;
	    \For{j := 1 \textbf{to} b3}
	    {
	    y := rand.range(0, length(ch.genes)-1)\tcp*{reverse y-th gene in}
	    ch[i].genes[y] := (ch[i].genes[y] + 1) \textbf{mod} 2
	    \tcp*{i-th chromosome}
	    }
    	\textit{nextGenChromosomes}.append(ch)\tcp*{update the list}
	}
	sort(\textit{nextGenChromosomes}) \tcp*{\textbf{Sort} the chromosomes by \textit{expenditure}}
	\KwReturn{\textit{nextGenChromosomes}}
	\caption{Mutation function for the New Population of the GA}\label{alg:NewPop1}
\end{algorithm}